\documentclass[final,1p, sort&compress]{elsarticle}
\usepackage{dsfont}
\usepackage{graphicx, amsmath, algorithm, algorithmic, slashbox}
\usepackage{multirow, subfig}
\usepackage[percent]{overpic}

\renewcommand{\Re}{\mathrm{Re}}
\renewcommand{\Im}{\mathrm{Im}}

\newcommand{\C}{\mathrm{C}}
\newcommand{\I}{\mathrm{I}}
\newcommand{\R}{\mathrm{R}}
\begin{document}
\begin{frontmatter}




\title{On the use of rational-function fitting methods for the solution of 2D Laplace boundary-value problems}
\author{Amit Hochman, Yehuda Leviatan, and Jacob K. White}
\begin{abstract}
A computational scheme for solving 2D Laplace boundary-value problems using rational functions as the basis functions is described. The scheme belongs to the class of desingularized methods, for which the location of singularities and testing points is a major issue that is addressed by the proposed scheme, in the context of the 2D Laplace equation. Well-established rational-function fitting techniques are used to set the poles, while residues are determined by enforcing the boundary conditions in the least-squares sense at the nodes of rational Gauss-Chebyshev quadrature rules. Numerical results show that errors approaching the machine epsilon can be obtained for sharp and almost sharp corners, nearly-touching boundaries, and almost-singular boundary data. We show various examples of these cases in which the method yields compact solutions, requiring fewer basis functions than the Nystr\"{o}m method, for the same accuracy. A scheme for solving fairly large-scale problems is also presented.
\end{abstract}
\end{frontmatter}

\section{Introduction}
Laplace boundary-value problems (BVPs) are prevalent in the applied sciences, and methods for their solution have been the subject of study for decades. Among these, prominent approaches are finite-difference and finite-element methods, integral-equation methods, and, in 2D, conformal-mapping techniques. While the Laplace equation is the simplest elliptic partial differential equation (PDE), improving methods for its solution is nevertheless of continued interest. At the focus of recent research are problems with small radii of curvature, nearly-touching boundaries, and singular, or almost-singular boundary data~\cite{helsing2008evaluation,helsing2008corner,driscoll2002schwarz,marshall2007convergence,bremer2010efficient,bremer2010universal}.

The technique described in this paper belongs to a class of methods known as desingularized methods, which are used for the solution of linear BVPs. What characterizes these methods is that the basis functions are elementary solutions of the homogenous PDE, and hence the singularities of the basis functions lie outside the BVP domain. The simplest example of a desingularized method is the method of images, which is only applicable in a handful of special cases for which the exact locations and amplitudes of the image sources are known analytically. For more general problems, an approximate solution can be obtained by setting the locations of the sources heuristically and solving for the amplitudes so that the boundary conditions are satisfied at a set of testing points. Various types of image sources and heuristics have been proposed over the years. Vekua~\cite{Vekua:1948} and Kupradze~\cite{kupradze1967approximate} are usually credited as the pioneers of these methods, and other early works include those of Fox, Henrici, and Moler~\cite{fox1967approximations}, Bergman~\cite{bergman1965numerical}, and Eisenstat~\cite{eisenstat1974rate}. Nowadays, the methods in this category go by various names, such as method of fundamental solutions~\cite{fairweather1998mfs,golberg1999method}, method of auxiliary sources~\cite{Kaklamani:2002,bucci2005nri}, generalized multipole technique~\cite{Hafner:1990}, method of fictitious sources~\cite[Ch. 1.5.6]{tayeb2004cfs,maystre2001smw}, and source-model technique~\cite{leviatan1988gff,boag1991aoe,hochman2007eas}.

Desingularized methods share many of the properties of integral-equation methods, and indeed can be viewed as such when the sources are distributed on a curve lying outside of the BVP domain. The basis functions used for the density functions are then Dirac delta functions. As the potentials due to these basis functions are smooth on the boundaries, it is customary to enforce the boundary conditions at a set of testing points. This obviates laborious panel integrals that slow down conventional Galerkin schemes for boundary integral equations, making the cost of forming the system matrix similar to that of a Nystr\"{o}m method. Also, the representation of the solution with elementary sources is very convenient for post-processing tasks such as calculating capacitances, charges, forces, etc., and the potential itself at arbitrary points.

\begin{figure}[h]
\centering
\subfloat[ ]{\scalebox{0.7}{\includegraphics{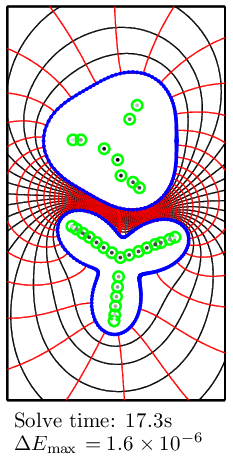}}}\hspace{0.5cm}
\subfloat[ ]{\scalebox{0.7}{\includegraphics{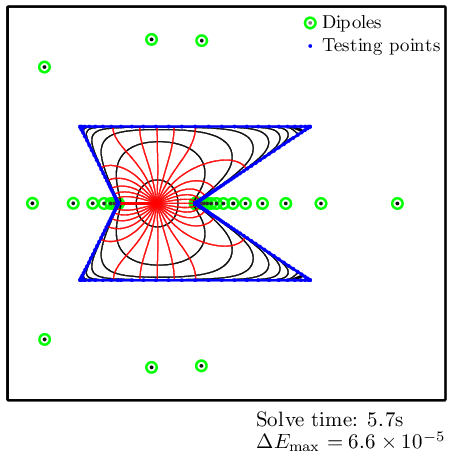}}}
\caption{Solutions obtained with a small number of basis functions, for (a) exterior and (b) interior Laplace Dirichlet problems. The number of dipoles is 30 in (a) and 60 in (b). Darker dots correspond to larger dipole amplitudes. } \label{fig:motivation}
\end{figure}

\begin{figure}[h]
\centering
{\scalebox{0.7}{\includegraphics{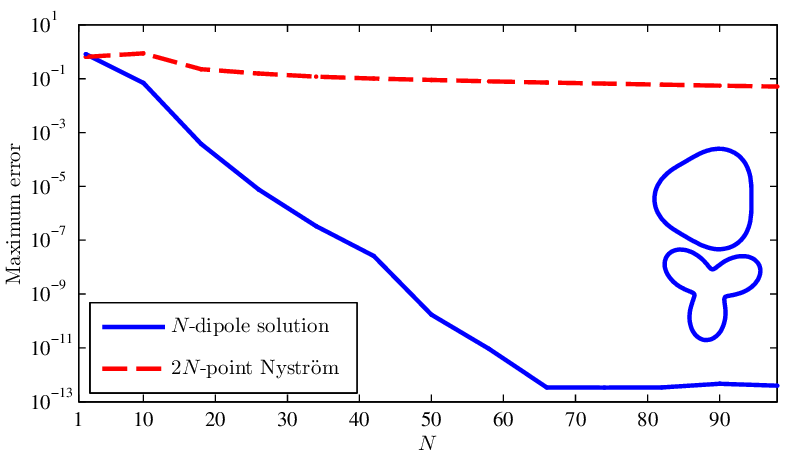}}}
\caption{For the problem shown in Fig.~\ref{fig:motivation}a, maximum errors on the order of $10^{-12}$ are obtained with no more than 70 dipoles. Using the Nystr\"{o}m method with the same number of basis functions yields errors of about 5\%.  } \label{fig:motivation_conv}
\end{figure}

Owing to the above properties, desingularized methods are among the simplest methods for solving linear BVPs. They can also be very accurate, provided the sources and testing points are located ``adequately". To substantiate this claim, we show a couple of results obtained with the method proposed in this paper. In Fig.~\ref{fig:motivation}a, 30 dipoles are used to solve the Laplace equation outside two closely spaced curves.
The solution is required to be $-1$ on the lower curve and $+1$ on the upper curve, and this boundary condition is satisfied with a maximum error of $\Delta E_\mathrm{max}= 1.7 \times 10^{-6}$. In Fig.~\ref{fig:motivation}b, an interior Dirichlet problem with two perfectly sharp reentrant corners is solved using 60 dipoles. The boundary condition is due to a monopole located inside the polygon and the maximum error is $\Delta E_\mathrm{max} = 2.5\times 10^{-5}$. The Nystr\"{o}m method, which is spectrally accurate when the boundaries are smooth~\cite{atkinson1997numerical}, is the standard against which the proposed method is compared with in this paper. Such a comparison, shown in Fig.~\ref{fig:motivation_conv}, reveals that when the boundary curves nearly touch, a desingularized method can indeed be much more accurate than the Nystr\"{o}m method.

The adequate location of the sources and testing points is a major unresolved issue with desingularized methods, and much work has been devoted to its analysis (\cite{leviatan1990analytic,katsurada1994charge,barnett2008stability,Chiba2009869,alves2009choice} are just a few references out of a large body of work). Source location affects both the rate of error decay and the effect of round-off errors which may limit the highest accuracy achievable. Also, the effective application of the fast multipole method (FMM)~\cite{greengard1987fast} appears to depend strongly on the source location~\cite{liu2005fast}. The problem with applying the FMM to a desingularized method is that the linear systems to be solved are generally ill-conditioned. In~\cite{liu2005fast}, the authors proposed a block-diagonal preconditioner that accelerates convergence, provided the distance between the sources curve and the boundary curve is less than a problem dependent critical distance.

Source location schemes that guarantee exponential convergence were developed in~\cite{katsurada1994charge} and~\cite{barnett2008stability,kangro2010convergence}, for Laplace and Helmholtz Dirichlet BVPs in simply-connected analytic domains, respectively. In these works, sources are distributed uniformly (in so-called conformal angle) on curves conformal with the boundary, and the resulting error in the solution is shown to decrease exponentially, in exact arithmetic. But placing sources on a conformal curve may be restrictive, and any a-priori source restriction requires a-priori knowledge of singularity locations.

A common approach has been to formulate the source-location problem as an optimization problem. If the number of sources is set in advance, finding the best source locations is a nonlinear least-squares problem, and various attempts have been made to use nonlinear optimization techniques to determine the source locations. Unfortunately, this optimization problem is not convex and has many local minima. Therefore, descent-type methods such as Gauss-Newton and Levenberg-Marquardt are sensitive to the initial source locations~\cite{fairweather1998mfs,mathon1977approximate,nishimura1984numerical}, while more robust methods that have been tried, such as simulated annealing~\cite{cisilino2002application} and genetic algorithms~\cite{heretakis2002analysis}, may take a very long time to converge. Moreover, if the positions of the sources are adapted, the positions of the testing points must be adapted as well. If this is not done, and the sources approach the boundary, the error at the testing points will not be representative of the error on the boundary. Alternatively, if the cost function of the optimization is defined as an integral on the squared error (as in~\cite{nishimura1984numerical}, for example), the quadrature for this integral must be adapted along with the source positions. This issue of adapting the testing points has been largely ignored in the literature; in this paper it is addressed by use of rational Gauss-Chebyshev quadrature rules. Perhaps the most practical approaches use geometry-based heuristics~\cite{eisler1989aoe,Hafner:1990,moreno2002mmm,barnett2008stability}. While these can be more than adequate in many engineering applications, they fail to take full advantage of the flexibility in locating the sources and their accuracy may be limited.

In this paper we describe a new method for determining appropriate locations for the sources and testing points. The method is tailored to the 2D Laplace equation, for which the solution can be represented as the real part of a complex potential which we approximate by the sum of a rational function and (for multiply-connected domains) a few logarithmic terms. We use well-established techniques for finding rational-function approximations to the boundary data, and the poles of these rational functions correspond then to the positions of dipole sources.

The major difficulty in applying this idea is that, in a Dirichlet problem, the boundary data corresponds to the real part of the complex potential on the boundary. The imaginary part is determined, up to an additive constant, by the Dirichlet-to-Neumann map. Of course, applying this map is tantamount to solving the Dirichlet problem. Similarly, in a Neumann problem, only the imaginary part of the complex potential on the boundary is known. The solution we propose for this problem is to estimate the missing information (say, the imaginary part in a Dirichlet problem) by solving the BVP with a given set of poles and then using this information to improve the pole locations. Assuming the pole locations are indeed improved, they can be used to find a better estimate of the imaginary part, which in turn is used to improve their locations further. We proceed in this way until the pole positions appear to have converged, the error drops below a preset tolerance, or a maximum number of iterations is reached.

Clearly, the process just described is not guaranteed to converge to the solution that would be obtained were the imaginary part known in advance. It is possible that the estimated imaginary part will be inaccurate, and the poles found based on this estimate will not yield any improvement on the next iteration. When this happens, the improvement in accuracy stalls. It is then usually, though not always, possible to increase the accuracy further by increasing the number of poles. In numerical experiments, we have found that despite this difficulty, errors approaching the machine epsilon can be obtained for various challenging configurations.

\section{Dirichlet problem, simply-connected domain}
We begin by considering the simplest case, in which we seek a real valued potential, $U(x,y)$, which obeys the Laplace equation. More precisely,
\begin{equation}
\nabla^2U(x,y) =0, \quad \text{for all } (x,y) \in \Omega,
\end{equation}
where $\Omega$ is a simply connected domain in $\mathds{R}^2$ bounded by a curve $\Gamma$, and $U$ must also obey a Dirichlet boundary condition,
\begin{equation}\label{eq:dirichlet}
    {\left. U \right|_{{\text{on }}{\Gamma}}} = f.
\end{equation}
The method we propose for solving this problem is to approximate the solution as a weighted combination of dipole potentials, and to use rational function fitting algorithms to determine the dipole locations.

\subsection{Dipole representation}
As is well-known, the potential $U$ can be sought as the real part of a complex function $W(z)$, where $z= x + iy$ and $W(z)$ is holomorphic and single valued in $\Omega$ (it is single valued because $\Omega$ is simply connected). The classical single- and double-layer potentials,
\begin{equation}\label{eq:singleLayer}
    W_\text{single}(z) = \int_\Gamma \sigma(z')\log(z' - z)dz',
\end{equation}
and
\begin{equation}\label{eq:doubleLayer}
    W_\text{double}(z) = \dfrac{1}{2\pi i}\int_\Gamma \dfrac{\mu(z')}{z' - z}dz',
\end{equation}
where $\sigma(z)$ and $\mu(z)$ are referred to as the monopole and dipole densities, respectively, have been used extensively to formulate integral equations~\cite{mikhlin1957integral,atkinson1997numerical,greenbaum1993laplace}. For example, a second-kind Fredholm equation is readily obtained by combining the double-layer potential,~\eqref{eq:doubleLayer}, with the Dirichlet boundary condition,~\eqref{eq:dirichlet}, or by combining the single-layer potential,~\eqref{eq:singleLayer}, with the Neumann boundary condition.

In desingularized methods, the most common approach is to use the single-layer potential,
\begin{equation}\label{eq:singleLayerDesing}
    W_{\Gamma'}(z) = \int_{\Gamma'} \sigma(z')\log(z' - z)dz',
\end{equation}
where $\Gamma'$ encloses, but does not intersect, $\Gamma$. Enforcing the boundary condition on $\Gamma$ leads to an ill-conditioned Fredholm equation of the first kind. Nevertheless, as discussed in~\cite{barnett2008stability}, if $W(z)$ can be continued analytically throughout the region enclosed between $\Gamma$ and $\Gamma'$, highly accurate solutions can be obtained, despite this ill-conditioning. The density $\sigma(z)$ is often approximated as a weighted combination of Dirac delta functions, as in
\begin{equation}\label{eq:diracDeltas}
    \sigma(z) = \sum_n \sigma_n \delta(z_n-z),
\end{equation}
in which case
\begin{equation}\label{eq:singleLayerApprox}
    W_{\Gamma'}(z) \approx \sum_n \sigma_n\log(z_n - z).
\end{equation}
Using a double-layer distribution instead of the single-layer distribution in~\eqref{eq:singleLayerDesing} does not yield a second-kind integral equation, as it does in the classical formulations, but this is nevertheless a viable option~\cite{barnett2010exponentially,Hafner:1990}. The approach in this work is similar, in that we also use dipole potentials as basis functions. However, the source points are not restricted to lie on any given curve, but may be placed anywhere outside $\Omega$. Assuming that $W(z)$ is also holomorphic for $z \in \Gamma$, it can always be expressed as a series of complex dipole potentials,
\begin{equation}\label{eq:ratFun}
    W(z) = \sum_{n=1}^{\infty} \dfrac{a_n}{z'_n - z}, \quad \forall z \in \Omega,
\end{equation}
where the $z'_n$ are referred to as the poles of $W(z)$.

According to a theorem by Wolff~\cite{wolffRatFun} (see also~\cite[Ch.1, Theorem 13]{walsh1965interpolation}), given the appropriate choice of poles, $z'_n \notin \overline{\Omega}$,~\eqref{eq:ratFun} converges uniformly on any closed set in $\Omega$. In this sense, we may say that the set of single-pole functions is a complete set, which can be used to approximate any function $W(z)$ which is holomorphic for $z \in \overline{\Omega}$. This theorem does not cover the case of $W(z)$ having a branch-point on $\Gamma$. Nevertheless, in Section~\ref{sssec:corners} we show a solution with error $< 10^{-10}$ even for such a case. The convergence theory of rational function approximations to functions with branch-points is rather involved, and we therefore refer the interested reader to a couple of relevant works~\cite{stahl1997convergence,borcea2006rational}.

Our algorithm is based on approximating the complex potential using $N$ poles plus a constant, or equivalently as an $N^\text{th}$ order rational function
\begin{equation}\label{eq:approxWithConst}
    W(z) \approx a_0 + \sum_{n=1}^N \dfrac{a_n}{z'_n - z} = \dfrac{P_N(z)}{Q_N(z)}
\end{equation}
where $P_N$ and $Q_N$ are complex-coefficient polynomials of degree $N$.
To determine $a_0, a_1, \ldots, a_N$, or equivalently, the coefficients of $P_N(z)$, we start with an initial guess for the poles, and associate a complex basis function with each pole,
\begin{equation}\label{eq:psiOfs}
    \phi^\C_n(z) = \dfrac{1}{z'_n-z},
\end{equation}
with real and imaginary parts,
\begin{equation}\label{eq:realAndImagPsi}
    \phi^\R_n(z) = \Re\left[\phi^\C_n(z)\right], \quad \phi^\I_n(z) = \Im\left[\phi^\C_n(z)\right].
\end{equation}
Note that $\phi^\R_n(z)$ and $-\phi^\I_n(z)$ are real-valued basis functions that correspond to $x$- and $y$-directed dipole potentials.
In the special case of the constant term, the complex basis function becomes a real-valued basis function,
\begin{equation}\label{eq:constant}
    \phi^\C_0 = \phi^\R_0 = 1.
\end{equation}
Defining $U(z) = \Re[W(z)]$ and $V(z) = \Im[W(z)]$, and using \eqref{eq:psiOfs}-\eqref{eq:constant},
\begin{align}\label{eq:approxU}
    U(z) & \approx \widehat{U}(z) = \sum_{n = 0}^{N} a_n^\R \phi_n^\R(z) - \sum_{n = 1}^{N}a_n^\I \phi_n^\I(z),
\end{align}
and its harmonic conjugate,
\begin{align}\label{eq:approxV}
    V(z) & \approx \widehat{V}(z) = \sum_{n = 1}^{N} a_n^\R \phi_n^\I(z) + \sum_{n=1}^N a_n^\I \phi_n^\R(z),
\end{align}
where the basis function coefficients $a^\R_n \equiv \Re(a_n)$ and $a^\I_n \equiv \Im(a_n)$ are determined by minimizing an error metric on the boundary $\Gamma$. More precisely,
\begin{equation}\label{eq:defA}
\mathbf{a} \equiv [a_0^\R, a_1^\R + i a_1^\I, \ldots, a_N^\R + i a_N^\I]^T
\end{equation}
is determined by solving the minimization problem
\begin{equation}\label{eq:minNormResidues}
        \min_{\mathbf{a}} \left|\left|\widehat{U} - f\right|\right|_\Gamma \equiv   \min_{\mathbf{a}}  \left< \widehat{U}-f, \widehat{U}-f\right>_\Gamma^{\frac{1}{2}},
\end{equation}
where the norm, $||\cdot||_\Gamma$, is induced by a Hermitian inner-product, $\left<\cdot, \cdot\right>_\Gamma$, the definition of which is deferred to Section~\ref{sec:innerProd}. The minimization in~\eqref{eq:minNormResidues} is equivalent to requiring that the residual, $f-\widehat{U}$, be orthogonal to the span of the basis functions, or equivalently,
\begin{subequations}\label{eq:Galerkin}
\begin{align}
    \left<\phi^{\R}_n, f - \widehat{U}\right>_\Gamma & = 0, \quad n = 0,1, \ldots, N,\\
    \left<\phi^{\I}_n, f - \widehat{U}\right>_\Gamma & = 0, \quad n = 1, \ldots, N.
\end{align}
\end{subequations}
Expanding the inner products and using~\eqref{eq:approxU} leads to the normal equations,
\begin{equation}\label{eq:normalEquations}
\begin{bmatrix}
    \mathbf{\Phi}^{\R\R} & \mathbf{\Phi}^{\R\I}\\
    \mathbf{\Phi}^{\I\R} & \mathbf{\Phi}^{\I\I}\\
\end{bmatrix}
\begin{bmatrix}
    \mathbf{a}^\R\\
    \mathbf{a}^\I\\
\end{bmatrix} =
\begin{bmatrix}
    \mathbf{f}_\phi^\R\\
    \mathbf{f}_\phi^\I\\
\end{bmatrix},
\end{equation}
where $\mathbf{\Phi}^{\alpha \beta}_{mn} = \left<\phi^{\alpha}_m,\phi^{\beta}_n\right>_\Gamma$ and $(\mathbf{f}^{\alpha}_\phi)_m = \left<f, \phi^{\alpha}_m\right>_\Gamma$.

Finding the basis function coefficients by least-squares minimization is common in desingularized methods. What is less common is our use of dipole potentials as basis functions, and also our use of a continuous error metric, as in~\eqref{eq:minNormResidues}. In Section~\ref{sec:innerProd} we will define a Hermitian inner product, and show that under mild conditions, enforcing the residual orthogonality~\eqref{eq:Galerkin} is equivalent to enforcing the boundary conditions at pole-location-related quadrature points.

\subsection{Pole relocation}\label{sec:poleReloc}
In the next step of the algorithm, the poles are moved to improve the approximation of $W(z)$. This is done by fitting a rational function to the complex function
\begin{equation}\label{eq:compPotToApprox}
    \widehat{W}(z) = f(z) + i\widehat{V}(z),\quad z \in \Gamma
\end{equation}
where $\widehat{V}(z)$ in~\eqref{eq:approxV} is based on the current set of poles. To obtain a ``better'' set of poles, we seek a rational approximation to $\widehat{W}(z)$ in the form $P(z)/Q(z)$ where both $P(z)$ and $Q(z)$ are polynomials of degree $\leq N$, and then use the zeros of $Q(z)$ as the ``better'' poles.

This rational fitting is done by minimizing a norm induced by the same inner-product used in~\eqref{eq:Galerkin}. Ideally, we would want to minimize $||P/Q-\widehat{W}||_\Gamma$, but this is a nonlinear and non-convex problem. So, instead, we follow the commonly-used strategy of first scaling the residual $P/Q-\widehat{W}$ by $Q$, and then correcting this scaling by dividing by
\begin{equation}\label{eq:QHat}
\widehat{Q}(z) = \prod_{n=1}^{N}\left(z'_n-z\right),
\end{equation}
which is an approximation for $Q(z)$ based on the current set of poles. The minimization problem is then
\begin{equation}\label{eq:minProbIRF}
    \min_{P, Q} \left|\left|\dfrac{P - Q\widehat{W}}{\widehat{Q}}\right|\right|_\Gamma,
\end{equation}
where $P$ and $Q$ are represented with polynomial bases, denoted $p_n$ and $q_n$, $n=0, 1, \ldots, N$, respectively. To avoid the trivial solution, $Q(z)$ is assumed to be monic in the $q_n$ basis. The minimization in~\eqref{eq:minProbIRF} leads to the orthogonality conditions,
\begin{equation}\label{eq:Galerkin2}
    \left<\dfrac{p_n}{\widehat{Q}}, \dfrac{P - Q\widehat{W}}{\widehat{Q}}\right>_\Gamma =0, \quad     \left<\dfrac{q_n \widehat{W}}{\widehat{Q}}, \dfrac{P - Q\widehat{W}}{\widehat{Q}}\right>_\Gamma =0, \quad n=0,1, \ldots, N.\\
\end{equation}
The choice of polynomial bases determines the conditioning of~\eqref{eq:Galerkin2}, and, as the monomial basis is a particularly poor choice, other choices have been proposed~\cite{gustavsen1999rational,deschrijver2007orthonormal,coelho1999robust}.

The above approach to rational fitting has a long history, dating back to Levy~\cite{levy1959complex} and Sanathanan and Koerner~\cite{sanathanan1963transfer}, and its application to real functions is much older. Usually, however, the minimization problem in~\eqref{eq:minProbIRF} corresponds to fitting a rational function at a finite set of points on the imaginary axis or the unit circle. For that more common case, many implementations have been suggested, including vector fitting (VF), iterated rational fitting (IRF)~\cite{coelho1999robust,vasilyevequivalence}, and recently, robust rational least-squares~\cite{gonnet2011robust}. We use VF as the default method, but in some cases it becomes poorly conditioned and we switch to a variant of IRF, the details of which are given in~\ref{app:irf}.

The main difference between VF and IRF is the choice of basis polynomials. In VF, the choice is $p_n(z) \equiv q_n(z) \equiv \ell_n(z)$, where
\begin{equation}\label{eq:Lagrange}
    \ell_n(z) =
    \begin{cases}
    \widehat{Q}(z) & \mbox{if } n=0\\
    \dfrac{\widehat{Q}(z)}{\left(z'_n-z\right)}  & \mbox{if } 1 \le n \le N.
    \end{cases}
\end{equation}
For $n =1,2, \ldots, N$, the $\ell_n(z)$ are, up to a scaling factor, the Lagrange polynomials for interpolating at the $z'_n$. As these polynomials are zero at all poles but one, they usually lead to a fairly well-conditioned linear system. However, if two or more poles are very close, conditioning may deteriorate. If this happens, we switch to using IRF, in which two polynomial bases are derived. For the $p_n(z)$, we use polynomials orthonormal with respect to the inner-product
\begin{equation}
\left<u,v\right>_{\Gamma P} = \left<\dfrac{u}{\widehat{Q}},\dfrac{v}{\widehat{Q}}\right>_\Gamma,
\end{equation}
and for the $q_n(z)$ we use polynomials orthonormal with respect to the inner product
\begin{equation}
\left<u,v\right>_{\Gamma Q} = \left<\dfrac{\widehat{W}u}{\widehat{Q}},\dfrac{\widehat{W}v}{\widehat{Q}}\right>_\Gamma.
\end{equation}
As a result of this choice, the linear system to be solved is made-up of two blocks of orthonormal columns, one for $P$ and one for $Q$ (see~\ref{app:irf}).

Once $Q(z)$ is obtained, its roots are computed and they become the poles for the next iteration. Using these poles, $\widehat{U}(z)$, the approximate solution of~\eqref{eq:dirichlet}, is recomputed by solving~\eqref{eq:normalEquations} and then using~\eqref{eq:approxU}. Similarly, using~\eqref{eq:approxV}, $\widehat{V}(z)$ is recomputed, $\widehat{W}(z)$ in~\eqref{eq:compPotToApprox} is updated, and the poles are relocated again.

When computing the roots of $Q(z)$ we may encounter roots inside $\Omega$. Although we do not discard these poles, their corresponding basis functions are not used to form $\widehat{U}(z)$ and $\widehat{V}(z)$. Hence, they are tacitly excluded from~\eqref{eq:approxU}-\eqref{eq:normalEquations}. As the algorithm progresses, these poles usually migrate outside of $\Omega$. Any poles that remain in $\Omega$ when the algorithm stops are discarded.


\subsection{Initial guess for the poles}\label{ssec:initial}
The proposed algorithm is found to be quite insensitive to the initial pole locations. We set them using the following heuristic, but even random initial locations seem to do fine. The initial guess for the $z'_n$ is obtained by solving~\eqref{eq:Galerkin2} assuming $\widehat{W}(z) = f(z)$, which is purely real, $\widehat{Q}(z) = 1$, and then allowing $P$ and $Q$ to be of degree $2N$. Out of the $2N$ zeros of $Q$ we pick $N$ zeros that are outside $\Omega$, and, if there are not enough of these, we supplement them with zeros inside $\Omega$ so that we have $N$ zeros in total. These $N$ zeros are used as the poles for the first iteration. The justification for using $2N$ poles along with a purely real $\widehat{W}(z)$ is based on considering the fitting problem when $\Omega$ is a disk. For a disk, a purely real $\widehat{W}(z)$ can be constructed with $N$ poles inside $\Omega$ and $N$ poles outside $\Omega$. Using only the poles outside, we obtain a function that is holomorphic in the disk and its real part fits $f(z)$ on $\Gamma$.


\subsection{Algorithm summary}
The main steps of the algorithm are as follows:
\begin{algorithm}[h!]
\caption{Solve simply-connected Dirichlet problem}
\label{alg:main}
\begin{algorithmic}[1]
\STATE Set the $z'_n$ to the initial guess, as explained in Section~\ref{ssec:initial}.
\WHILE {stopping criterion not met}
\STATE Solve~\eqref{eq:normalEquations} for the $a^{\R}_n$ and $a^{\I}_n$. Use only $z'_n \notin \Omega$.
\STATE Use the $a^{\R}_n$ and $a^{\I}_n$ to form $\widehat{V}$ according to~\eqref{eq:approxV} and $\widehat{W}$ according to~\eqref{eq:compPotToApprox}.
\STATE Form $\widehat{Q}$ according to~\eqref{eq:QHat}, use all $z'_n$.
\STATE Solve~\eqref{eq:Galerkin2} for $P$ and $Q$.
\STATE $z'_n = \text{roots}(Q)$.
\ENDWHILE
\end{algorithmic}
\end{algorithm}

\section{Inner product definition and quadrature rules}\label{sec:innerProd}

In each iteration of Algorithm~1, two norm minimization problems must be solved: one to determine an approximate complex potential given the poles,~\eqref{eq:minNormResidues}, and one to update the poles using the approximate complex potential,~\eqref{eq:minProbIRF}.
Both problems can be formulated as follows: given a set basis functions $\psi_1, \psi_2, \ldots \psi_{N_\psi}$ defined on $\Gamma$, and a function to approximate, $g$, also defined on $\Gamma$, we wish to determine the vector of coefficients $\mathbf{c}$ that solves the norm minimization problem
\begin{equation}\label{eq:continuousLS}
   \min_{\mathbf{c}} \left\|\sum_{n=1}^{N_\psi} c_n\psi_n - g\right\|_\Gamma.
\end{equation}
The unique solution to this problem is given by
\begin{equation}\label{eq:pinv}
    \mathbf{c} = \mathbf{\Psi}^{-1}\mathbf{g_\psi},
\end{equation}
where $\mathbf{\Psi}_{mn} = \left<\psi_m,\psi_n\right>_\Gamma$ and $(\mathbf{g_\psi})_m = \left<g,\psi_m\right>_\Gamma$. It can be easily shown that $\mathbf{\Psi}$ is Hermitian positive definite and hence invertible, provided that the $\psi_n$ are linearly independent (that this is so for the $\phi_n^\R$ and $\phi_n^\I$ follows from a trivial modification of Theorem 4 in~\cite{kupradze1967approximate}, where it was proved in a 3D context).
The inner product used in this work is
\begin{equation}\label{eq:ip}
    \left<u,v\right>_\Gamma = \int_0^1 u\left(z(s)\right)v^*\left(z(s)\right)\lambda(s)ds,
\end{equation}
where $\lambda(s)$ denotes a positive weight function, and $z(s)$ is a parametrization of $\Gamma$. In view of this definition, the inner-product integrals in $\mathbf{\Psi}$ and $\mathbf{g_\psi}$ may be evaluated using a quadrature rule. If~\eqref{eq:ip} can be evaluated with a single quadrature rule for all $u$'s and $v$'s occurring in inner-product integrals in $\mathbf{\Psi}$ and $\mathbf{g_\psi}$, it is simpler and more numerically stable not to use~\eqref{eq:pinv} explicitly, but instead to sample $\sqrt{\lambda(s)}g(s)$ and the $\sqrt{\lambda(s)}\psi_n(s)$ at the quadrature nodes, and to set-up an overdetermined linear system for $\mathbf{c}$,
\begin{equation}\label{eq:discreteLS}
\begin{bmatrix}
        \sqrt{\lambda_1}\psi_1(s_1) & \cdots & \sqrt{\lambda_1}\psi_{N_\psi}(s_1)\\
        \vdots & \ddots & \vdots\\
        \sqrt{\lambda_{K_\psi}}\psi_1(s_{K_\psi}) & \cdots & \sqrt{\lambda_{K_\psi}}\psi_{N_\psi}(s_{K_\psi})\\
\end{bmatrix}
\begin{bmatrix}
    c_1\\
    \vdots\\
    c_{N_\psi}\\
\end{bmatrix} \approx
\begin{bmatrix}
    \sqrt{\lambda_1}g(s_1)\\
    \vdots\\
    \sqrt{\lambda_{K_\psi}}g(s_{K_\psi})\\
\end{bmatrix},
\end{equation}
where $s_1, s_2, \ldots, s_{K_\psi}$ are the quadrature nodes and $\lambda_1, \lambda_2, \ldots, \lambda_{K_\psi}$ are the quadrature weights. Equation~\eqref{eq:discreteLS} may also be written as
\begin{equation}\label{eq:LSLambda}
    \mathbf{\Lambda}^{\frac{1}{2}}\mathbf{\widehat{\Psi}} \mathbf{c} \approx \mathbf{\Lambda}^{\frac{1}{2}}\mathbf{\widehat{g}},
\end{equation}
where
\begin{gather}
    \mathbf{\Lambda}  = \mbox{diag}(\lambda_1, \lambda_2, \ldots, \lambda_{K_\psi}) \nonumber  \\
    \mathbf{\widehat{g}}  = [g(s_1), g(s_2), \ldots, g(s_{K_\psi})]^T  \\
    \mathbf{\widehat{\Psi}}_{mn} = \psi_n(s_m). \nonumber
\end{gather}
The least-squares solution to~\eqref{eq:LSLambda} is given analytically by
\begin{equation}\label{eq:LSAnalytic}
\mathbf{c} = \left(\mathbf{\widehat{\Psi}}^H \mathbf{\Lambda} \mathbf{\widehat{\Psi}}\right)^{-1}\mathbf{\widehat{\Psi}}^H \mathbf{\Lambda\widehat{g}},
\end{equation}
where superscript $H$ denotes Hermitian transpose. It is more numerically stable, however, to obtain this solution by QR decomposition with column pivoting applied to~\eqref{eq:discreteLS}. If the quadrature rule is exact, then the solution given in~\eqref{eq:LSAnalytic} also solves~\eqref{eq:continuousLS} (see, e.g.,~\cite[Ch. 4]{atkinson1997numerical}).

\subsection{Quadrature rules for~\eqref{eq:minNormResidues}}
The basis functions used to solve~\eqref{eq:minNormResidues}, restricted to $\Gamma$ and seen as a function of the parameter $s$, are the real and imaginary parts of
\begin{equation}\label{eq:psiK}
    \phi^\C_n(s) = \dfrac{1}{z'_n-z(s)}.
\end{equation}
Clearly, constructing quadrature rules that will integrate inner products of these functions for arbitrary $z(s)$ is not practical. However, the parametrization of $z(s)$ is ours to choose, and we choose it to be a piecewise rational function, i.e.,
\begin{equation}\label{eq:piecewisepoly}
    z(s) = \begin{cases}
    r_1(s) & \text{if } 0 \leq s < h_1\\
    r_2(s) & \text{if } h_1 \leq s < h_2\\
    \vdots\\
    r_{M}(s) & \text{if } h_{M-1} \leq s < 1\\
    \end{cases},
\end{equation}
where the $r_m(s)$ are rational functions in $s$. This choice of parametrization is quite general, as piecewise rational functions can be used to represent any geometry of practical interest, and they are easily generated from $\Gamma$ using readily available libraries (we use both Matlab's spline toolbox and the Chebfun system~\cite{chebfunv4}). With this choice of parametrization, the $\phi^\C_n(s)$ are themselves piecewise-rational functions, each piece having known poles in the complex $s$ plane. Quadrature rules for products of rational functions with known poles are readily available, and this is the motivation for our choice of parametrization\footnote{Another advantage of this parametrization is that determining whether a pole is inside or outside of $\Omega$ is easy (see~\ref{app:in_or_out}).}. However, the actual basis functions used in~\eqref{eq:minNormResidues} are $\phi^\R_n(s)$ and $\phi^\I_n(s)$, and these are not piecewise-rational functions of $s$, as they depend on both $s$ and $s^*$. To overcome this difficulty, we use the Schwartz reflection principle. We write the real basis functions as
\begin{equation}\label{eq:poleresidue}
    \phi^\R_n(s) = \dfrac{1}{2}\begin{cases}
    \dfrac{1}{z'_n-r_1(s)} + \dfrac{1}{\left[z'_n-r_1(s)\right]^*} & \text{if } h_0 \leq s < h_1\\[3ex]
    \dfrac{1}{z'_n-r_2(s)} + \dfrac{1}{\left[z'_n-r_2(s)\right]^*} & \text{if } h_1 \leq s < h_2\\
     & \!\!\!\!\vdots\\
    \dfrac{1}{z'_n-r_{M}(s)} + \dfrac{1}{\left[z'_n-r_{M}(s)\right]^*} & \text{if } h_{M-1} \leq s < h_{M}    \end{cases},
\end{equation}
where $h_0 = 0, \ h_{M}=1$. Note that the expression for $\phi^\I_n$ is similar. Although $\phi^\R_n$ and $\phi^\I_n$ are functions of both $s$ and $s^*$, for real $s$ they can be written as piecewise rational functions in $s$ alone. These rational functions can be obtained by replacing $\left[z'_n-r_m(s)\right]^*$ by $\left[z'_n-r_m(s^*)\right]^*$ everywhere in~\eqref{eq:poleresidue}, and similarly for $\phi^\I_n$. We then have,
\begin{equation}\label{eq:multiIndex}
    \phi^{(\R,\I)}_n(s) = \dfrac{{\displaystyle \prod\limits_{l=1}^{L_m}\left(s-\beta^{(\R,\I)}_{lmn}\right)}}{{\displaystyle \prod\limits_{l=1}^{L_m}\left(s-\alpha_{lmn}\right)\left(s-\alpha^*_{lmn}\right)}} \quad \text{for } s \in [h_m, h_{m+1}), \quad m=0, 1, \ldots, M-1,
\end{equation}
where $L_m$ is the degree of the numerator or denominator of $r_m(s)$, whichever is higher. In~\eqref{eq:multiIndex}, $\alpha_{lmn}$ is the $l^\text{th}$ pole associated with the $n^\text{th}$ basis function, restricted to the $m^\text{th}$ section of the boundary, i.e., it is the $l^\text{th}$ zero of $z'_n-r_m(s)$.

Quadrature rules that are exact for the inner products of the basis functions are readily obtained using rational Gauss-Chebyshev quadrature rules as building blocks. These rules integrate exactly integrals of the form
\begin{equation}\label{eq:ratInt}
    \int_{-1}^{1}{R(s)T^*(s^*)}\dfrac{1}{\sqrt{1-s^2}}ds,
\end{equation}
where $R(s)$ and $T(s)$ are rational functions that have poles at known locations (and need not be proper)~\cite{van2008algorithm}. Once the $\alpha_{lmn}$ are determined, a quadrature rule for each section of the boundary is calculated, in $O(N L_m)$ time, using the Matlab toolbox Rcheb~\cite{van2008algorithm}. The rules are then concatenated to obtain a single rule for the entire boundary.


Using these quadrature rules implies a specific choice for the inner-product weight function, denoted $\lambda(s)$ in~\eqref{eq:ip}, namely,
\begin{equation}\label{eq:lambda}
    \lambda(s) = \begin{cases}
    \lambda_1(s) & \text{if } h_0 \leq s < h_1\\
    \lambda_2(s) & \text{if } h_1 \leq s < h_2\\
    \vdots\\
    \lambda_{M}(s) & \text{if } h_{M-1} \leq s < h_{M}\\
    \end{cases},
\end{equation}
where
\begin{equation}\label{eq:scaledLambda}
    \lambda_m(s) = \dfrac{h_{m+1}-h_m}{\sqrt{(h_{m+1}-h_m)^2-\left(2s - h_{m+1}-h_m\right)^2}}.
\end{equation}
This choice of $\lambda(s)$ is dictated primarily by the ease with which rational Gauss-Chebyshev quadrature rules may be computed. However, it also has the advantage that when the boundary consists of a single rational-function section, using the Chebyshev weight function implies that the approximation will be close to optimal in the max-norm sense, provided the boundary data is smooth enough~\cite{cheney1982introduction,van2008algorithm}. When the boundary is composed of more than one section, this is no longer so, but the Chebyshev weight function still helps suppress Runge's phenomenon near $s=0$ and $s=1$. Lastly, this choice of weight function makes our error metric dependent on the parametrization $z(s)$. To avoid this, one could choose $\lambda(s) = |dz/ds|$, which would make the inner product a line integral and hence invariant to the choice of parametrization, but this choice would complicate the construction of the quadrature rules. A compromise is to make the weight of each section of the boundary proportional to its length, and this can be done with the chosen weight function by making the size of the interval $[h_m, h_m+1]$ proportional to the length of the $m^\text{th}$ section of the boundary.

We still need to ensure that the inner-products between basis functions and the boundary data $f\left(z(s)\right)$ are calculated accurately. Assuming the boundary data is smooth on each section of the boundary, this can be accomplished by making the quadrature rules exact for improper rational functions with a sufficiently high numerator polynomial degree. In Rcheb, this is done by adding poles at infinity when generating the quadrature rules.

\subsection{Quadrature rules for problem~\eqref{eq:minProbIRF}}\label{ssec:quadForIRF}
The inner products for this second norm minimization problem are given in~\eqref{eq:Galerkin2}. All the integrands in this case can be written as rational functions of $s$, with poles due to the zeros of $\widehat{Q}\left(z(s)\right)$ as well as the poles of $\widehat{W}\left(z(s)\right)$.
Recalling the definition of $\widehat{W}\left(z(s)\right)$ given in~\eqref{eq:compPotToApprox}, its real part is just the boundary data, and hence is assumed piecewise smooth, while the imaginary part has the same poles as those used for~\eqref{eq:minNormResidues}. To these poles we must add all zeros of $\widehat{Q}\left(z(s)\right)$, which are the zeros of all the $z'_n-r_m(s)$, including any poles $z'_n \in \Omega$. Although the quadrature rules are calculated quickly by Rcheb, it is usually more economical to call it only once, with the poles for~\eqref{eq:minProbIRF} and then to solve both equations with the same set of quadrature points.

The proposed scheme for evaluating inner-product integrals with quadrature yields linear matrix equations very much like those of conventional desingularized methods, except that quadrature dictates the location and weight of each testing point. This scheme takes into account the location of the sources, the shape of the boundary, and the variation of the boundary data, and minimizes a continuous error norm.

The total number of testing points used, $K$, is given by
\begin{equation}\label{eq:ntesting}
    K = M (N_\infty +1) + \left(3N_\mathrm{out} + N_\mathrm{in} \right)\sum_{m=1}^{M}L_m,
\end{equation}
with the notation

\vspace{10pt}
\begin{tabular}{ll}
$M$ & number of rational-function sections of the boundary\\
$N_\infty$ & number of infinite poles added\\
$N_\mathrm{out}$ & number of poles outside of $\Omega$\\
$N_\mathrm{in}$ & number of poles inside of $\Omega$\\
$L_m$ & degree of numerator or denominator of $r_m(s)$, whichever higher.
\end{tabular}
\vspace{10pt}

\noindent As the scheme is based on Gaussian quadrature, this number is in a sense optimal for computing with the continuous error norm, and the computation of the testing points is done with a fast and accurate library routine. We have found that, in all but the simplest cases, many uniformly distributed testing points are needed to match the accuracy obtained by the Gaussian quadrature scheme, so the extra computational burden is generally worthwhile. However, the representation of the boundary with piecewise rational functions should be made as simple as possible, i.e., with the fewest sections and lowest orders. If the boundary is broken into many sections, then taking into account all the poles when computing the quadrature rule for each section can be excessive. It is better to take into account only poles that are close to a given section. The rest of the poles can be dealt with more efficiently by a small number of poles at infinity. We refer to this process as pole pruning.

\subsubsection{Pole pruning}\label{sec:pruning}
\begin{figure}[h]
\centering {\scalebox{1}{\includegraphics{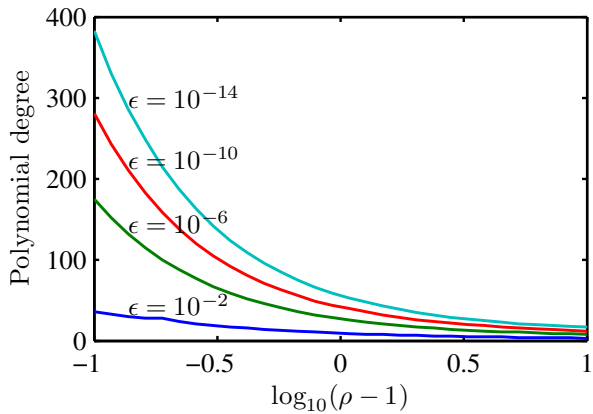}}}
\caption{Polynomial degree that yields approximations of $R_{s_0}(s)$ with errors less than $\epsilon$, as a function of $\rho$. For $\rho \approx 10$, errors under $10^{-14}$ can be obtained with polynomial degrees under 20.} \label{fig:chebyErrs}
\end{figure}

Adding $N_\infty$ poles at infinity makes the quadrature rule exact for integrands that are products of a rational function and a polynomial of degree $2N_\infty-1$. Hence, we can omit poles that are distant enough that their corresponding potentials are smooth. To determine which poles are far enough, we need to estimate the error committed in approximating a rational function by a polynomial on the integration interval. As a prototypical rational function we take a third-order pole function,
\begin{equation}\label{eq:typicalRat}
    R_{s_0}(s) = \dfrac{1}{(s-s_0)^3},
\end{equation}
as this is the highest pole order that will appear in the integrands, and estimate the error committed in approximating it by a polynomial, as a function of the pole $s_0$. The max-norm error of the polynomial that interpolates at Chebyshev nodes is almost minimal and it is the same for all $s_0$ on the ellipse
\begin{equation}\label{eq:paramEllipse}
    s(\phi) = \dfrac{\rho e^{i\phi} + \rho^{-1} e^{-i\phi}}{2}, \quad \phi \in [0, 2 \pi],
\end{equation}
where $\rho>1$ is the so-called \emph{elliptical radius} (see, e.g.,~\cite{trefethen2009approximation}). For any $s_0 \notin [-1, 1]$ we can calculate the elliptical radius corresponding to the ellipse which passes through it,
\begin{equation}\label{eq:rho_of_s}
    \rho(s_0) = \max\left(\left|s_0 + \sqrt{s_0^2-1}\right|, \left|s_0 - \sqrt{s_0^2-1}\right|\right).
\end{equation}
Although there are bounds for the error in approximating a function such as $R_{s_0}(s)$ by a polynomial, we found it more practical to tabulate the degrees of polynomials that yield approximation errors less than $\epsilon$, for relevant values of $\epsilon$ and $\rho(s_0)$ (Fig.~\ref{fig:chebyErrs}).
To obtain these polynomial degrees, we again used the Chebfun system, which truncates a Chebyshev expansion once the expansion coefficients drop sufficiently below $\epsilon$. The tabulation is then interpolated or extrapolated as necessary so that we can rapidly calculate the optimal $\rho$ beyond which poles can be discarded and replaced by poles at infinity, without incurring errors much larger than $\epsilon$. As shown in Fig.~\ref{fig:chebyErrs}, for all but the most distant poles, the choice of $\epsilon$ influences the required polynomial degree appreciably. Hence, if only moderate accuracy is sought, significant reduction may be gained by using a larger $\epsilon$. To take advantage of this, we set $\epsilon$ according to
\begin{equation}\label{eq:epsilon}
    \epsilon = \min(10^{-4}, 0.01\Delta E),
\end{equation}
where $\Delta E$ is the normalized error obtained when solving~\eqref{eq:Galerkin} in the previous iteration (a more precise definition is given below in~\eqref{eq:deltaE}). Thus, if the solution is not very accurate, fewer quadrature nodes are used, but their number is increased if the algorithm is allowed to iterate and higher accuracy is achieved. In the very first iterations, $\Delta E$ can be large, so we bound epsilon from above by $10^{-4}$. Of course, this rule is heuristic and has been derived from numerical experiments in which it appeared to do well, but it is by no means optimal and a different rule could perform better.

\subsubsection{Froissart doublets}
Admittedly, the quadrature scheme just described is somewhat complicated, but we have found this complication to be worthwhile. The reason is related to the occurrence of pole-zero pairs, called Froissart doublets~\cite{froissart1969approximation,gilewicz1999pade,pachón2009piecewise}, which almost coincide with the boundary. These pole-zero pairs are a numerical artifact, and they do not contribute to the fidelity of the fit, but instead they fit errors in the data. In our case, the errors in the imaginary part of the data are quite large in the first iterations, so these artifacts occur often. If uniformly distributed testing points are used, the doublets frequently occur closer to the boundary than the separation between testing points. When the poles of these doublets are used to solve~\eqref{eq:minNormResidues}, the resulting potential is inaccurate between testing points.
In turn, the estimate of $\Im[W(z)]$ is inaccurate, and the improvement in accuracy from one iteration to the next may stall. One could try to filter these poles in various ways, for example, based on their small residues and distances from the boundary, but this would limit the obtainable accuracy. By using the proposed quadrature scheme, all the poles outside of $\Omega$, no matter how close they are to its boundary, can be safely used to solve~\eqref{eq:minNormResidues}. 

\begin{figure}[h]
\centering
\subfloat[ ]{\rotatebox{0}{\scalebox{0.6}{\includegraphics{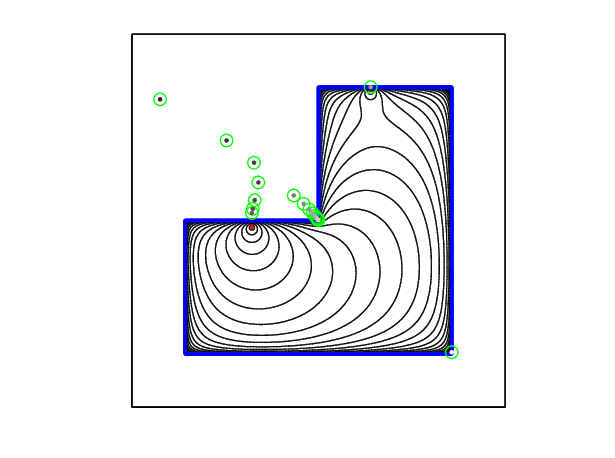}}}}
\subfloat[ ]{\rotatebox{0}{\scalebox{0.6}{\includegraphics{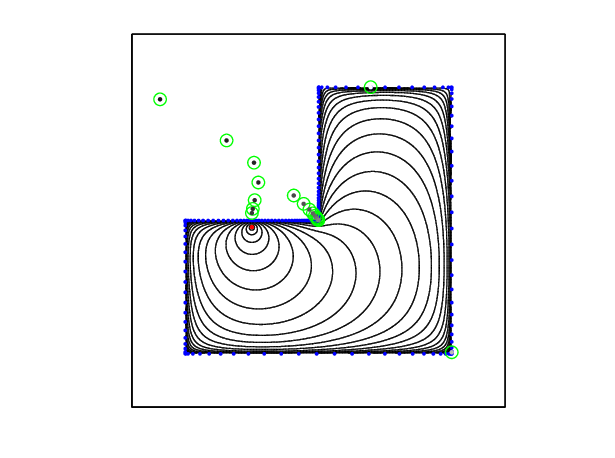}}}}
\caption{Equipotential lines for a monopole (marked by a red dot) inside an L-shaped region. In (a), 2000 points were uniformly distributed on $\Gamma$, in (b), 203 points were distributed by the quadrature scheme.} \label{fig:quadDemo}
\end{figure}

In Fig.~\ref{fig:quadDemo}a, an artifact caused by a Froissart doublet can be observed near the top of $\Gamma$. The pole of the doublet is extremely close to the boundary (the distance is about $10^{-9}$ of the perimeter of $\Gamma$), which explains the appearance of the artifact despite having used 2000 uniformly distributed testing points. When the quadrature scheme is used, with just 203 points, the artifact disappears (Fig.~\ref{fig:quadDemo}b).

\section{Dirichlet problem, multiply-connected domain}\label{sec:exterior}
Other Laplace BVPs can also be solved with the proposed scheme, after appropriate adaptations. As an example of a Dirichlet problem in a multiply-connected domain we consider the exterior Dirichlet problem; multiply-connected interior problems can be treated similarly. In the exterior Dirichlet problem, the potential $U(x,y)$ obeys the Laplace equation everywhere outside of a set of disjoint domains $\Omega_1, \Omega_2, \ldots, \Omega_J$ bounded by contours $\Gamma_1, \Gamma_2, \ldots, \Gamma_J$, respectively. As before, Dirichlet boundary conditions are imposed,
\begin{equation}\label{eq:exteriorBC}
     U|_{\mathrm{on} \ \Gamma_j} = f_j, \quad j = 1,2,\ldots J\\
\end{equation}
and in addition, the potential must be bounded at infinity. Of course, the poles must now be restricted to lie inside the $\Omega_j$. Also, the complex potential $W(z)$ is no longer single-valued, but can be written as the sum of $J$ logarithmic terms and a single-valued holomorphic function $\widetilde{W}(z)$ (see, e.g.,~\cite[\S{29}]{mikhlin1957integral}). We have,
\begin{equation}\label{eq:wtilde}
   W(z) = \widetilde{W}(z) + \sum_{j=1}^{J}A_j \log\left(z-z^\Omega_j\right), \quad z^\Omega_j \in \Omega_j,
\end{equation}
where the $A_j$ may be assumed real, and the $z^\Omega_j$ can be anywhere inside $\Omega_j$, preferably as far from its boundary as possible. If the $A_j$ were known, the logarithmic terms could be subtracted from the boundary data, and the algorithm for the simply-connected Dirichlet problem could be applied to $\widetilde{W}(z)$. Our approach is therefore to estimate the $A_j$ iteratively, along with the imaginary part of $\widetilde{W}(z)$. On each iteration, we solve
\begin{equation}
\begin{split}\label{eq:exteriorDirichlet}
 &   \sum_{n = 0}^{N_\mathrm{out}} a_n^\R \phi_n^\R(z) - \sum_{n = 1}^{N_\mathrm{out}}a_n^\I \phi_n^\I(z) + \sum_{j=1}^{J}A_j \log\left|z-z^\Omega_j\right| \approx f(z), \\[1ex]
 &   \text{with } f(z) = f_j(z) \quad \text{for } z \in \Gamma_j,\\
 &   \mathrm{subject \ to \ } \sum_{j=1}^{J}A_j = 0,
\end{split}
\end{equation}
for the $a_n^\R$, the $a_n^\I$, and the $A_j$. The condition that the sum of $A_j$'s is zero, needed to ensure the solution is bounded at infinity, can be enforced by eliminating one of the $A_j$, say, $A_{J} = - \sum_{j=1}^{J-1}A_j$. Instead of~\eqref{eq:exteriorDirichlet} we then have,
\begin{equation}\label{eq:exteriorDirichlet2}
\sum_{n = 0}^{N_\mathrm{out}} a_n^\R \phi_n^\R(z) - \sum_{n = 1}^{N_\mathrm{out}}a_n^\I \phi_n^\I(z) + \sum_{j=1}^{J-1}A_j \log\left|\dfrac{z-z^\Omega_j}{z-z^\Omega_{J}}\right| \approx f(z).
\end{equation}
As before, we solve~\eqref{eq:exteriorDirichlet2} by sampling at quadrature nodes, weighting the equation by the square roots of the quadrature weights, and solving the linear system in the least-squares sense. In contrast to the previous case, however, the inner products involving the logarithmic basis functions will not be integrated exactly. We must therefore add poles at infinity when deriving the quadrature rules for each section of the boundary. Fortunately, the logarithmic singularity is weak, so when the $z^\Omega_j$ are not too close to the boundary, the restriction of the logarithmic term to the boundary curve can be approximated well by piecewise polynomials. The poles at infinity added by the pole-pruning process of Section~\ref{sec:pruning} usually suffice, although extra poles at infinity can be added if necessary.

Once~\eqref{eq:exteriorDirichlet2} has been solved, a rational function is fitted to an estimate of the boundary values of $\widetilde{W}(z)$,
\begin{equation}\label{eq:wh}
    \widetilde{W}(z) \approx f(z) - \sum_{j=1}^{J-1}A_j \log\left|\dfrac{z-z^\Omega_j}{z-z^\Omega_{J}}\right| + i\widehat{V}(z), \quad z \in \bigcup_{j=1}^{J}{\Gamma_j} = \Gamma.
\end{equation}
The real part is estimated as $f(z)$, the known values of $\Re[W(z)]$ on the boundaries, with the real part of the logarithmic terms subtracted. The estimate for $\Im[\widetilde{W}(z)]$ is $\widehat{V}(z)$, given by~\eqref{eq:approxV}, and it remains as it was for the simply-connected problem when $\widetilde{W}(z)$ and $W(z)$ were the same.

\section{Exterior Neumann problem}
Now consider the case when the normal derivative of the potential on the boundary is prescribed. For brevity, we discuss only the exterior problem, which can be formulated as
\begin{equation}
\begin{split}\label{eq:neumann}
 &   \nabla^2U(x,y) = 0, \quad \text{for all } (x,y) \notin \bigcup_{j=1}^{J}\Omega_j,\\
 &   {\left. \dfrac{\partial U}{\partial n} \right|_{{\text{on }}{\Gamma_j}}} = f_j,\\
 & \lim_{|z|\rightarrow \infty}{\!\!U(x,y)|z|} < \infty, \quad z = x+iy,
\end{split}
\end{equation}
where the $\Omega_j$ are again simple disjoint domains bounded by the $\Gamma_j$. This problem is uniquely solvable if and only if
\begin{equation}\label{eq:neumannCondition}
\sum_{j=1}^{J}{\int_{\Gamma_j}{f_j(z)|dz|}} = 0.
\end{equation}

The approach we adopt for this problem follows~\cite[\S{32}]{mikhlin1957integral}. As in the exterior Dirichlet problem, we seek the solution as the real part of a function $W(z)$ that is holomorphic outside the $\Omega_j$, and which can be written as a sum of a single-valued part, $\widetilde{W}(z)$, and $J$ logarithmic terms. In this case, however, the logarithmic terms are known in advance:
\begin{equation}\label{eq:logterm}
A_j = -\dfrac{1}{2 \pi}\int_{\Gamma_j}f_j(z)|dz|,
\end{equation}
and they can be subtracted from the boundary data. After this subtraction,
\begin{equation}\label{eq:neumannSubtractLogs}
    \widetilde{f}_j(z) = f_j(z) - \sum_{j=1}^{J}A_j \log\left|z-z^J_j\right|,
\end{equation}
can be used to write a Neumann boundary condition for $\Re\left[\widetilde{W}(z)\right]$. Denoting $\widetilde{W}(z) = \widetilde{U}(z) + i \widetilde{V}(z)$, this condition reads,
\begin{equation}\label{eq:tildeNeumann}
\left.\dfrac{\partial \widetilde{U}}{\partial n}\right|_{\text{on }\Gamma_j} = \tilde{f}_j.
\end{equation}
If $\widetilde{W}(z)$ is approximated by a weighted combination of dipole potentials, it becomes inconvenient to enforce~\eqref{eq:tildeNeumann} directly, as the differentiation increases the order of the poles. Instead, we integrate~\eqref{eq:tildeNeumann} with respect to arc-length and use the Cauchy-Riemann equations to obtain,
\begin{equation}\label{eq:vtilde}
\widetilde{V}(z) = \int_{z_j}^z{ \tilde{f}_j(z)|dz|} + v_j \equiv F_j(z) + v_j,
\end{equation}
where the integration is along $\Gamma_j$, from an arbitrary point $z_j$ to the point $z$, and the $v_j$ are constants yet to be determined. Finding a harmonic function $\widetilde{V}$ that satisfies the boundary condition~\eqref{eq:vtilde} is known as a modified Dirichlet~\cite{mikhlin1957integral} or Muskhelishvili~\cite{pogorzelski1966integral} problem, and given the condition at infinity in~\eqref{eq:neumann}, it is solvable only for a unique set of $v_j$ constants. These can be determined along with an approximate solution for  $\widetilde{V}$, by solving
\begin{equation}\label{eq:neumannResidues}
\sum_{n = 1}^{N_\mathrm{out}} a_n^\R \phi_n^\I(z) + a_n^\I \phi_n^\R(z) - v_j \approx F_j(z), \quad \text{for} \ z \in \Gamma_j,
\end{equation}
for the $a^\R_n$, the $a^\I_n$, and the $v_j$.
As before,~\eqref{eq:neumannResidues} is solved in the least-squares sense after sampling the basis functions at quadrature nodes and weighting the equations by the square roots of quadrature weights. Once~\eqref{eq:neumannResidues} is solved, the rational fitting algorithm is applied to an estimate of the boundary values of $\widetilde{W}(z)$ given by,
\begin{equation}\label{eq:whNeumann}
    \widetilde{W}(z) \approx  \sum_{n = 1}^{N_\mathrm{out}} a_n^\I \phi_n^\I(z) + a_n^\R \phi_n^\R(z) + i \left[ F_j(z) + v_j\right], \quad \text{for} \ z \in \Gamma_j.
\end{equation}
As expected, in the Neumann problem, it is the \emph{imaginary} part of the complex potential that is prescribed by the boundary data (up to $J$ additive constants), while its real part and the $J$ constants are estimated iteratively. Note that the $n=0$ term corresponding to the constant basis function is omitted from the sum in~\eqref{eq:whNeumann} so as to conform with the boundary condition at infinity in~\eqref{eq:neumann}.

The formulation just described differs from one that is commonly used for Neumann problems in integral-equation formulations (e.g., in~\cite{greenbaum1993laplace}). These formulations typically use a single-layer potential so that enforcing the boundary condition of~\eqref{eq:neumann} leads to a second-kind Fredholm equation. We could have formulated our solution similarly, i.e., by representing the potential as a weighted combination of monopole potentials. Enforcing the boundary condition would then lead to a weighted combination of dipole potentials approximating the normal derivative of $U$. The chosen formulation does have some advantages, however. First, the functions to be approximated, the $F_j(s)$, are smoother than the $f_j(s)$, so fewer basis functions may be needed for their approximation. Second, a common situation in Neumann problems is that the $A_j$ of~\eqref{eq:logterm} are zero, so that the complex potential $W(z)$ is single valued. This occurs when the boundary is composed of a single contour, or when the $\Omega_j$ represent impenetrable objects. In these cases, it is convenient to approximate $W(z)$ using single-valued basis functions, instead of having a superposition of non-single-valued functions that can be made single-valued by an appropriate choice of branch cuts, as would occur if monopoles were used. For instance, using the proposed formulation, it is easy to obtain equipotential contours and field lines as the level sets of $\Re\left[W(z)\right]$ and $\Im\left[W(z)\right]$, without having to keep track of the proper Riemann-sheet on which to evaluate the logarithmic kernel.

\section{Large-scale problems}\label{sec:large}
The proposed algorithm for setting the poles and testing-points works well for moderately-sized geometries, but its computational cost is too high for larger ones. However, it is still possible to tackle large-scale problems by partitioning the geometry into smaller domains and running the algorithm for each of these domains separately, possibly in parallel. Once the poles and testing points are obtained for the entire geometry, we solve the full problem with the locally-optimized poles and testing points. If the cost of solving the full problem becomes prohibitive, it may be possible to use an iterative domain-decomposition approach, along the lines of~\cite{stupfel1996fast}, \cite{lee2005non}, or \cite{geuzaine2010amplitude}, which could also be used to improve the locations of the poles and testing points from one iteration to the next.

\begin{figure}[h]
\centering \rotatebox{270}{\scalebox{0.4}{\includegraphics{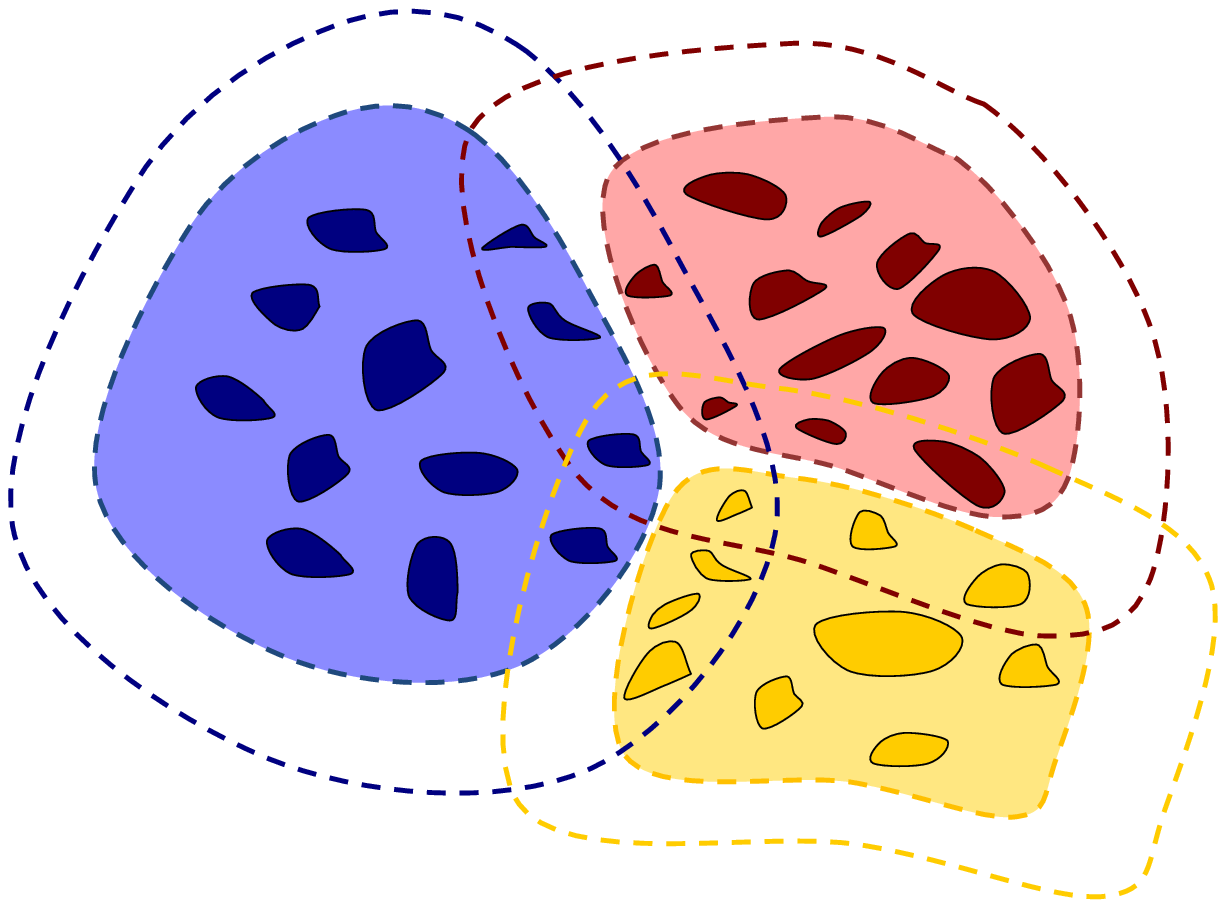}}}
\caption{Boundary curves partitioned into three clusters with buffer regions (delimited by dashed lines).} \label{fig:clust}
\end{figure}

We propose a scheme for geometries made up of a large number of simple closed curves. The scheme consists of partitioning the boundary curves into clusters so that each curve belongs to a single cluster, as shown in Fig.~\ref{fig:clust}.
Various algorithms for obtaining such a partition exist, and we use the simplest one, the $k$-means clustering, or Lloyd's, algorithm~\cite{lloyd1982least}. To keep the partitioning simple, each curve is entirely within one cluster and represented by its centroid, on which the clustering algorithm operates. Then, to each cluster we add a ``buffer region,'' as shown in Fig.~\ref{fig:clust}, so that curves close to the edge of the cluster have all their nearby interactions taken into account. The buffer region is obtained by adding to each curve all the curves with centroids closer than some radius, and this radius is chosen as twice the distance to the nearest centroid among all other curves. Poles that, after the local optimization, are found inside the buffer region are not used when solving the full problem.
\section{Numerical results} \label{sec:results}
The proposed method was implemented in Matlab, and results for a number of sample problems were obtained. This implementation, which is freely available for download~\cite{download}, was run on a 2GHz Intel Xeon machine with 8GB of RAM. The error metric adopted in this work is $\Delta E$,
\begin{equation}\label{eq:deltaE}
    \Delta E = \dfrac{\|\mathbf{\hat{u}} - \mathbf{f}\|_w}{\|\mathbf{f}\|_w},
\end{equation}
where $\mathbf{f}$ and $\mathbf{\hat{u}}$ are vectors of samples at the quadrature points of the exact and approximate boundary values, respectively. The $\|\cdot\|_w$ norm includes multiplication by the quadrature weights, so that it is an approximation of the continuous norm. The residual, $\widehat{U} -U$, is itself a harmonic function throughout $\overline{\Omega}$ and hence is largest at the boundary. We also used the maximum-norm error,
\begin{equation}\label{eq:deltaEmax}
    \Delta E_\mathrm{max} = \dfrac{\underset{z \in \Gamma}{\max}\left|f(z) - \widehat{U}(z)\right|}{{||f(z)||_{L^2(\Gamma)}/||1||_{L^2(\Gamma)}}},
\end{equation}
where $||1||_{L^2(\Gamma)}$ is just the length of $\Gamma$. In practice, we estimated $\Delta E_\mathrm{max}$ by sampling $f$ and $ \widehat{U}$ on a large number of points on the boundary.

\subsection{Interior Dirichlet Problem}
Problems of increasing difficulty were considered. Where applicable, results were compared to results obtained with the standard Nystr\"{o}m method, implemented along the lines of~\cite{greenbaum1993laplace}. However, we have not compared with more sophisticated Nystr\"{o}m schemes, such as those of~\cite{helsing2008corner,helsing2008evaluation,bremer2010efficient,bremer2010universal}, which have been recently proposed for handling more ``difficult'' geometries and excitations.

\subsubsection{A few well-separated poles}
We begin with the simplest problems, in which the exact solution is known in advance to have just a few, well-separated, poles, i.e.,
\begin{equation}\label{eq:simpleF}
    f(z) = \Re \left[\sum_{k=1}^{N_0}{\dfrac{a^0_k}{z'_k - z}}\right], \quad z'_k \notin \Omega, \quad z \in \Gamma
\end{equation}
Although this excitation suits our basis functions perfectly, the solution obtained is not always very accurate even if $N=N_0$. If the estimate of $\Im\left[W(z)\right]$ is inaccurate enough, the poles may be relocated to poor positions, and using these poles will not lead to a more accurate estimate of $\Im\left[W(z)\right]$. This problem can be alleviated by using $N>N_0$, though the number of poles that must be added beyond $N_0$ depends on the location of the poles of the solution and also on the shape of the boundary. This dependence is explored below.

\begin{figure}[h]
\centering \rotatebox{0}{\scalebox{0.8}{\includegraphics{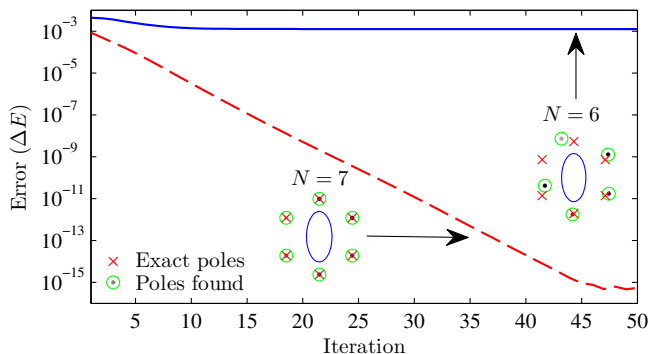}}}
\caption{Error vs. iteration for an ellipse and 6-pole boundary data. One of the poles remains inside the ellipse,  both for $N=6$ and $N=7$.} \label{fig:ellipse}
\end{figure}

A plot of error vs. iteration count is shown in Fig.~\ref{fig:ellipse}, for an ellipse with unit semi-major axis and an aspect ratio of two, and boundary data as in~\eqref{eq:simpleF} with $N_0 = 6, a^0_k = k/N_0$, and $z'_k = 1.5\exp\left[i\pi (2k+1)/6\right]$.
When $N=N_0$, the error does not decrease much below 0.1\%, but when $N=N_0+1$, the error drops exponentially, approaching the machine epsilon after 45 iterations. The reason for this dramatic difference in accuracy is that when $N=N_0$, a single pole remains inside $\Omega$ and the estimate of $\Im[W(z)]$ obtained with the remaining five poles is not accurate enough to make it migrate outside of $\Omega$. Consequently, the improvement in accuracy stalls. In contrast, when $N=N_0+1$, the extra pole allows for a better approximation of the imaginary part and six poles appear outside of $\Omega$. They are then relocated until the error approaches the machine epsilon.

A natural possibility is to increase $N$ adaptively, whenever the improvement in accuracy stalls. For example, one can calculate $\sigma_K(\Delta E)$, the standard deviation of $\Delta E$ in the previous $K$ iterations, and add a pole whenever
\begin{equation}\label{eq:addSource}
    \sigma_K(\Delta E) < \mu_K(\Delta E) \epsilon,
\end{equation}
where $\mu_K(\Delta E)$ is the $K$-average of $\Delta E$ and $\epsilon$ is a small number. A more sophisticated strategy might be found along the lines of~\cite{pachón2009piecewise}, which could also be used to decrease the number of poles if necessary. We have implemented the simple strategy with $K=5$ and $\epsilon=0.5\%$, which appears to yield a practical tradeoff between the conflicting goals of only adding poles when necessary, and keeping the number of iterations moderate. This adaptive scheme, which makes the algorithm fully automatic, is available as part of the downloadable package~\cite{download}. However, to keep the results section concise, we only show results for the nonadaptive version.
%
%
%

\begin{figure}[h]
\centering \rotatebox{0}{\scalebox{0.5}{\includegraphics{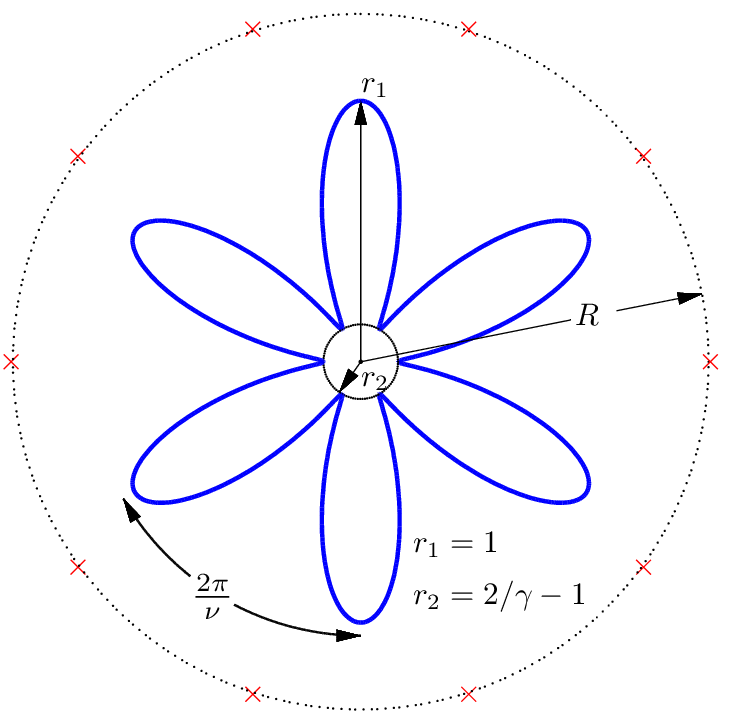}}}
\caption{Boundary curve and poles of boundary data used in Table~\ref{tab:add}.} \label{fig:trigpolyGeom}
\end{figure}

We consider next boundary curves of the form,
\begin{equation}\label{eq:trigpoly}
    z(s) = \left[1/\gamma + (1/\gamma-1)\cos(2 \pi \nu s)\right] e^{j2 \pi s},\quad s \in [0,1]
\end{equation}
with $\gamma = 1.75$, and $\nu = 2,4,6.$
A plot of the $\nu = 6$ curve is shown in Fig.~\ref{fig:trigpolyGeom}, together with a set of ten points distributed on a circle of radius $R$, which represent the poles of the boundary data.
This radius was set to 1.1, 1.5, and 3, while the number of poles was kept to ten. The algorithm was run for a maximum of 100 iterations and we determined the smallest $N$ required so that $\Delta E < 10^{-14}$ when the algorithm stops. The results of this experiment are summarized in Table~\ref{tab:add} where this value of $N$ is given, together with $\Delta E_\mathrm{max}$, estimated by evaluating the approximate solution at 20000 boundary points.
\begin{table}[h]
\caption{Results for boundary curves as in Fig.~\ref{fig:trigpolyGeom}. \label{tab:add}}
\centering
\resizebox{0.75\columnwidth}{!}{
\begin{tabular}{|c|c|c|c|c|c|c|}
\cline{1-7}
\multirow{2}{*}{\backslashbox{$R$}{$\nu$}} & \multicolumn{2}{|c|}{2} & \multicolumn{2}{|c|}{4} & \multicolumn{2}{|c|}{6}\\ \cline{2-7}
 & $\Delta E_{\max}$ & $N$ & $\Delta E_{\max}$ & $N$ & $\Delta E_{\max}$ & $N$ \\ \cline{1-7}
1.1 & $8.8\times 10^{-15}$ & $10$ & $1.4\times 10^{-14}$ & 11 & $1.6\times 10^{-14}$ & 14 \\
1.5 & $3.1\times 10^{-15}$ & $18$ & $5.6\times 10^{-15}$ & $22$ & $8.3\times 10^{-15}$ & $18$ \\
3 & $4.3\times 10^{-15}$ & $25$ & $7.8\times 10^{-15}$ & $27$ & $1.4\times 10^{-14}$ & $29$ \\
\hline
\end{tabular}}
\end{table}
In all cases, $\Delta E_{\max}$ is of the same order of magnitude as the tolerance for $\Delta E$, although in a few cases it is slightly larger than $10^{-14}$. We see that, in contrast to conventional desingularized- and integral-equation- methods, the proposed method works best when the poles of $f$ are close to the boundary, adding only a few spurious poles during the solution. Evidently, when the poles are further away their locations are more difficult to ascertain. We note that $\nu$ has only minor influence on $N$.

\subsubsection{Almost-singular boundary data}
\begin{figure}[h]
\centering
\includegraphics[width=3.5in]{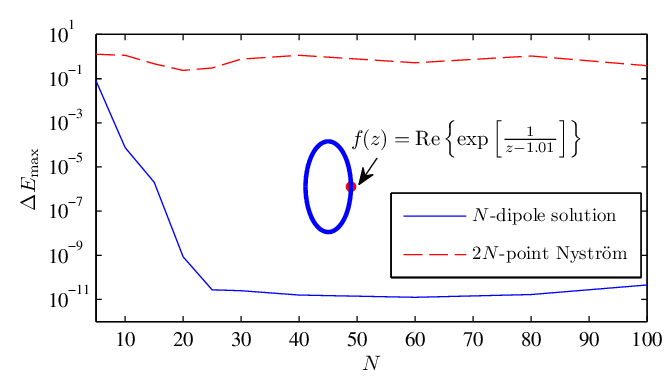}
\caption{Almost-singular boundary data. Comparison of the error vs. number of basis functions ($2N$) obtained with the proposed method and the Nystr\"{o}m method.} \label{fig:essential}
\end{figure}

When the boundary data is singular, or almost so, the accuracy of most numerical methods suffers. A possible remedy is to add basis functions that can approximate the singular boundary data to the set of basis functions being used (for an example in the context of desingularized methods, see~\cite{alves2006crack}). In the proposed method this happens automatically, as the basis functions are chosen from a large over-complete set. In the next example, the boundary is an ellipse with unit semi-minor axis, and an aspect ratio of two, and the boundary data is,
\begin{equation}\label{singularExp}
f(z) = \Re\left\{\exp\left[\dfrac{1}{z-1.01}\right]\right\}, \quad z\ \in \Gamma.
\end{equation}
This implies that the exact complex potential has an essential singularity very close to the boundary. Nevertheless, errors on the order of $\Delta E_\mathrm{max} = 10^{-11}$ are obtained with no more than 25 poles (Fig.~\ref{fig:essential}).
These errors could not be reduced further, and in fact, as more poles are added the error increases slightly. Compared to the proposed algorithm, the Nystr\"{o}m method yields very large errors, and its convergence rate is very slow. For 200 collocation points the Nystr\"{o}m method yields $\Delta E_\mathrm{max} = 0.38$, and, outside the range of the figure, 1000 collocation points still yield only $\Delta E = 0.05$.

Of course, for the same number of basis functions, the computation time of the Nystr\"{o}m method is shorter by a factor related to the number of pole relocation iterations. Even so, the Nystr\"{o}m method is more time-consuming in this example for any reasonable accuracy. For example, for $\Delta E = 0.05$, the Nystr\"{o}m method takes about 6s, while the proposed method takes 2.5s and it uses five poles. For higher accuracies the differences increase and it quickly becomes impractical to use a direct solver with the Nystr\"{o}m method. If the boundary data is smoother, for example, if the essential singularity is moved to $z=1.1$, the Nystr\"{o}m method can be faster. It requires 0.9s and $N=400$ for $\Delta E = 10^{-5}$, whereas the proposed method requires 2.5s and five poles to achieve the same accuracy.

At each point on the error curves in Fig.~\ref{fig:essential} we calculated the error of an $N$-pole solution and a $2N$-point Nystr\"{o}m solution, so as to match the number of basis functions. The maximum error for both the Nystr\"{o}m method and the proposed method was calculated by evaluating the approximate potential on $8N$ boundary points. To obtain this potential in the Nystr\"{o}m method we used a $8N \times 8N$ collocation matrix, which multiplies the double-layer density interpolated trigonometrically for its values at the $8N$ points.

\subsubsection{Almost-sharp corners}
\begin{figure}[h]
\centering
\raisebox{3.5cm}{
\begin{minipage}{2.25cm}
$N = 35$\\
$K = 343$\\
$\Delta E = 7 \times 10^{-5}$\\
Iterations: 35\\
Time: 5.1s
\end{minipage}
}
\rotatebox{0}{\scalebox{0.75}{\includegraphics{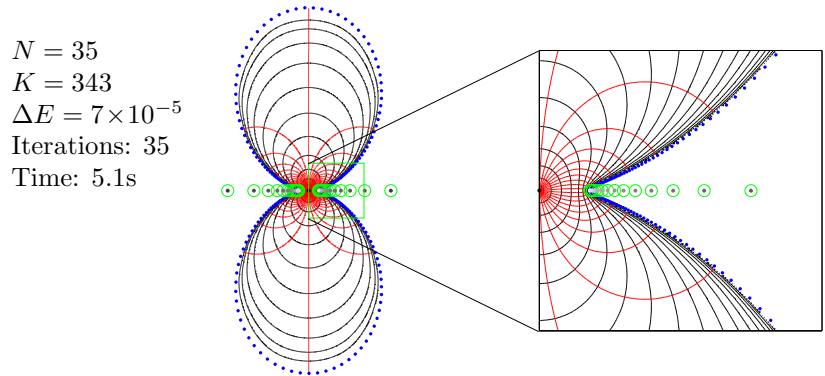}}}
\caption{Equipotential contours (logarithmically spaced) and field lines for a monopole inside the curve: \mbox{$z(s) = \left[1/1.9 + (1/1.9-1)\cos(4 \pi s)\right] e^{j2 \pi s}$}. The poles of the solution are marked by green circles, with centers colored according to the magnitude of the residues (darker means larger residues). The testing points are marked by blue dots. } \label{fig:trigpolyGreenfields}
\end{figure}

The influence of the shape of the boundary in the previous examples is limited to determining where the solution is sampled; it does not influence the location of poles of the solution as these are set by the boundary data. In the next example, we consider a Green's-function type problem, where a boundary curve of form~\eqref{eq:trigpoly} with $\nu = 2$ is excited by a monopole potential,
\begin{equation}\label{eq:ls}
    f(s) = \log|z(s)|.
\end{equation}
As the parameter $\gamma \rightarrow 2$ in~\eqref{eq:trigpoly} the problem becomes more difficult, because the minimal radius of curvature decreases, and also the sharp tips of the boundary approach the monopole and each other. A solution of this problem with $\gamma = 1.9$, obtained with 35 poles is shown in Fig.~\ref{fig:trigpolyGreenfields}. 
The inhomogeneous distribution of poles and testing points is clearly necessary in order to approximate the rapidly varying field where the radius of curvature is small. To calculate the testing points, the boundary was parameterized  by a single polynomial in $s$, of degree $34$, which approximates $f(s)$ to machine precision. Equipotential contours and field lines are also shown in Fig.~\ref{fig:trigpolyGreenfields}. These were obtained as the level sets of $\Re\left[W(z)\right]$ and $\Im\left[W(z)\right]$ by use of Matlab's `contour' command. The evaluation of $W(z)$ throughout $\Omega$ is trivial, as it only requires the summation of the partial-fraction expansion for $W(z)$. This may be contrasted with the more elaborate schemes needed to evaluate integral equation solutions throughout the domain, especially close to the boundary (see, e.g.,~\cite{helsing2008evaluation}).

\begin{figure}[h]
\centering \rotatebox{0}{\scalebox{0.95}{\includegraphics{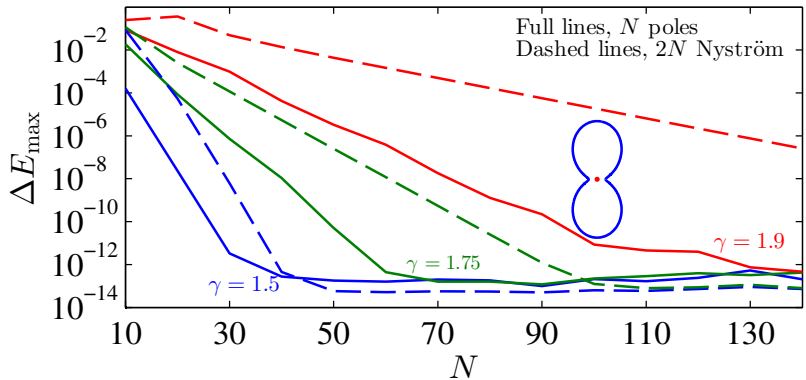}}}
\caption{Comparison of the proposed algorithm and the Nystr\"{o}m method for boundary curves with decreasing minimal radius of curvature. The difference between the methods is more significant for smaller minimal radius of curvature.} \label{fig:trigpolyComp}
\end{figure}

Plots of $\Delta E_\mathrm{max}$ vs. $N$ are shown in Fig.~\ref{fig:trigpolyComp} for curves with $\gamma = 1.5, 1.75, 1.9$, together with results obtained with the Nystr\"{o}m method.
The points for the Nystr\"{o}m method were distributed uniformly in the parameter $s$ of~\eqref{eq:trigpoly}, implying a higher density near the origin, which helps in resolving the rapid variation in the double-layer density there. The proposed algorithm was allowed to run for a maximum of 100 iterations, though the poles stopped moving after roughly 50 iterations. The rate of convergence with respect to the number of poles is approximately exponential, and the minimal error obtained is similar to that of the Nystr\"{o}m method. It is also evident that the proposed method generally yields a more accurate solution per basis function than the Nystr\"{o}m method, and that the difference between the two increases with $\gamma$.

\subsubsection{Perfectly sharp corners}\label{sssec:corners}
\begin{figure}[h]
\centering
\raisebox{3cm}{
\begin{minipage}{3.15cm}
$N = 30$\\
$M = 230$\\
$\Delta E_\text{max} = 1.8 \times 10^{-5}$\\
Iterations: 60\\
Time: 3.2s
\end{minipage}
}
\rotatebox{0}{\scalebox{0.85}{\includegraphics{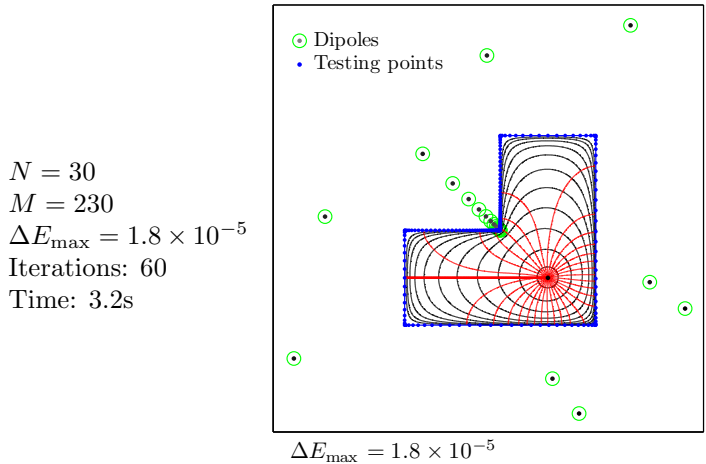}}}
\caption{Equipotential contours (logarithmically spaced) and field lines for a monopole inside an L-shaped domain.} \label{fig:corner}
\end{figure}

\begin{figure}[h]
\centering \rotatebox{0}{\scalebox{0.6}{\includegraphics{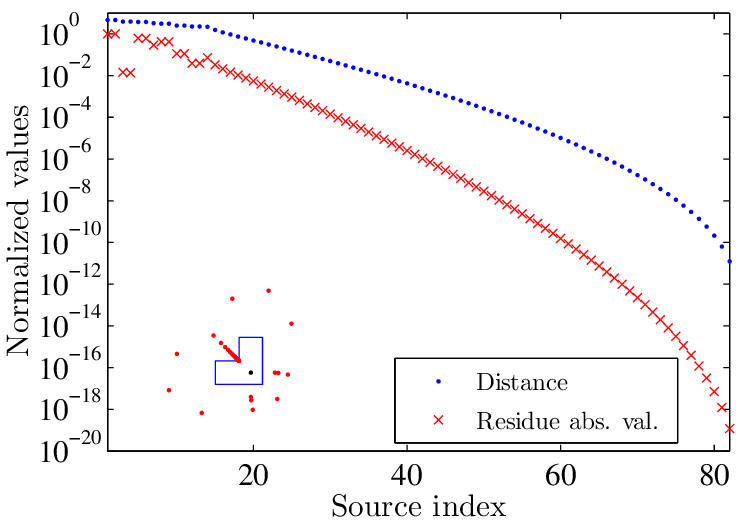}}}
\caption{Distances to reentrant corner, normalized with respect to the distance of the exciting line-source from the corner, and corresponding absolute value of residues, normalized with respect to the largest one.} \label{fig:distToCorner}
\end{figure}

When the boundary has sharp corners, the potential may have singularities at the corners, making it impossible to  represent with simple poles alone. It is still possible, however, to obtain very good approximate solutions with a moderate number of poles.
Such an example is shown in Fig.~\ref{fig:motivation}b, and another one is shown in Fig.~\ref{fig:corner}.
As can be observed in both figures, the poles and testing points cluster near the reentrant corners, and appear to be approximating a branch-cut which exists in the analytical continuation of the solution (see~\cite{borcea2006rational} for a discussion of the curves on which poles cluster when approximating algebraic functions by rational functions). The distribution of poles near the reentrant corner is examined in Fig.~\ref{fig:distToCorner}, where the poles are plotted in descending order of distance from the corner, for a solution with 82 poles of the L-shaped boundary.
The distances, which are normalized with respect to the distance of the exciting line-source from the corner, are seen to decrease at an almost constant exponential rate, with the closest poles clustered even more densely. The closest pole is placed at a normalized distance of about $10^{-11}$, indicating that poles must be located very precisely if high accuracy is to be achieved. Moreover, it seems inevitable that the testing points must also be adapted so that they are located with comparable precision. One might wonder whether the sharp turn in the equipotential lines near the corner is achieved with poles having large residues of differing phases. As shown in Fig.~\ref{fig:distToCorner}, this is not the case. Although the residues have absolute values that span 19 orders of magnitude, the smallest ones correspond to the closest poles.

\begin{figure}[h]
\centering
\subfloat[ ]{\rotatebox{0}{\scalebox{0.7}{\includegraphics{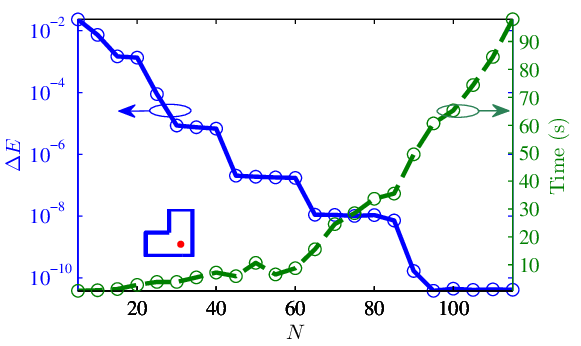}}}}
\subfloat[ ]{\rotatebox{0}{\scalebox{0.7}{\includegraphics{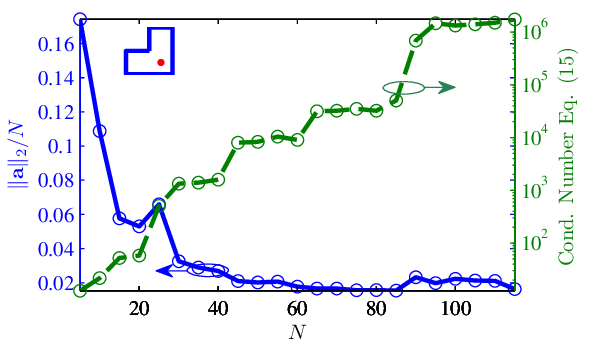}}}}
\caption{L-shaped domain, monopole excitation. Despite singularities on the boundary, errors smaller than $10^{-10}$ can be obtained with no more than 100 poles (a). Also, although the condition number increases with $N$, the norm of the vector of residues (normalized with respect to $N$) decreases with $N$ (b).} \label{fig:cornerConv}
\end{figure}

We now examine the convergence of the algorithm with $N$ (Fig.~\ref{fig:cornerConv}). As seen in Fig.~\ref{fig:cornerConv}a, the dependence of the error on $N$ is not as smoothly exponential as in the previous examples.
This more uneven error decrease is due to inaccuracies in the imaginary part that result in poles converging to locations inside $\Omega$. For example, when $N=20$ four of the poles remain inside $\Omega$, while when $N=30$ all poles move outside of $\Omega$. As a result, $\Delta E$ decreases by more than two orders of magnitude when going from $N=20$ to $N=30$. As shown in Fig.~\ref{fig:cornerConv}b, the condition number increases with $N$, but the norm of the vector of residues (normalized with respect to $N$) decreases. This is similar to~\cite{barnett2008stability}, in which such behavior was observed when the sources enclosed all the singularities of the analytic continuation of the potential. It is remarkable that this occurs in the present example as the potential cannot be continued analytically beyond the corners.

\subsection{Exterior Dirichlet problem}
\begin{figure}[h]
\centering
\subfloat[ ]{\includegraphics[width=2.55in]{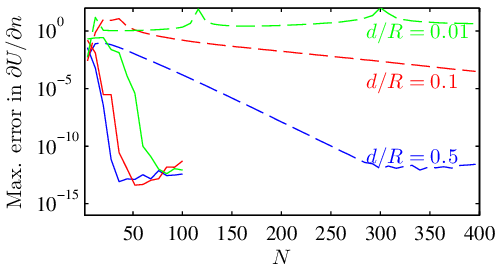}}\hspace{0.25cm}
\subfloat[ ]{\begin{overpic}[width=2.55in]{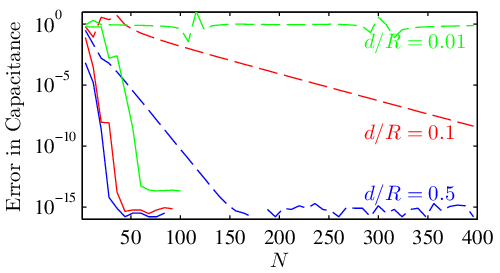}
    \put(48,13){\includegraphics[width=0.35in]{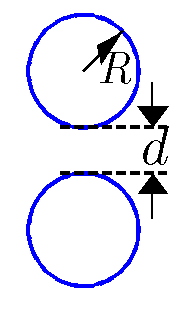}}
\end{overpic}}
\caption{Two closely-spaced curves. Full lines, $N$-pole solution. Dashed lines, $2N$ Nystr\"{o}m solution. The accuracy of the proposed method is far less sensitive to the separation between circles than the Nystr\"{o}m method.} \label{fig:twoCylinders}
\end{figure}

The exterior Dirichlet problem differs from the interior problem by the presence of logarithmic terms in the former. Also, the region in which the poles may be placed in the exterior problem is bounded, which probably makes their location easier. A typical difficulty in exterior problems occurs when the boundary is composed of two or more curves that nearly touch. An example of this sort consists of two circular curves with boundary values $-1$ on the lower curve and $+1$ on the upper one, and relative separations $d/R = 0.01, 0.1, 0.5$, as shown in Fig.~\ref{fig:twoCylinders}.
This example has an analytic solution~\cite{smythe1950static} to which we may compare the results. Specifically, seen as a two-conductor electrostatic problem, a global parameter of interest is the capacitance, proportional to the magnitude of the logarithmic terms, and a local parameter of interest is the charge density, proportional to the normal derivative of the potential. As can be seen in Fig.~\ref{fig:twoCylinders}, for both of these parameters, the proposed method requires far fewer degrees of freedom than the Nystr\"{o}m method, and this difference increases as the conductors approach. Also of note is that the proposed method is far less sensitive to the separation between conductors. To a good approximation, the $N$ needed to achieve a given error scales like $1/d$ when using the Nystr\"{o}m method, but it scales like $\log(1/d)$ when using the proposed method.

\subsection{Large-scale problems: exterior Neumann example}
In the last example, we solve a Neumann problem of potential flow around impenetrable ellipses of random radius, eccentricity, and orientation, many of which are very close to one another. This geometry was generated by a simple modification of the method described in~\cite{helsing2008evaluation} for generating a distribution of random circles. The velocity field far from the ellipses is assumed uniform, of unit amplitude, and in the $x$ direction. The corresponding BVP is then given by~\eqref{eq:neumann} with $f_j(z) = -\cos\left[\theta_j(z)\right]$ where $\theta_j(z)$ is the angle between the outward-pointing normal to the $j^\text{th}$ curve and the $x$ axis.

\begin{figure}[h]
\centering
\scalebox{0.8}{\includegraphics{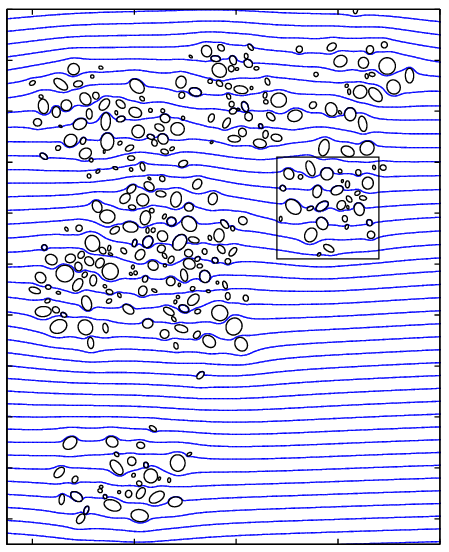}} \ \ \
\raisebox{0.8cm}{\scalebox{0.5}{\includegraphics{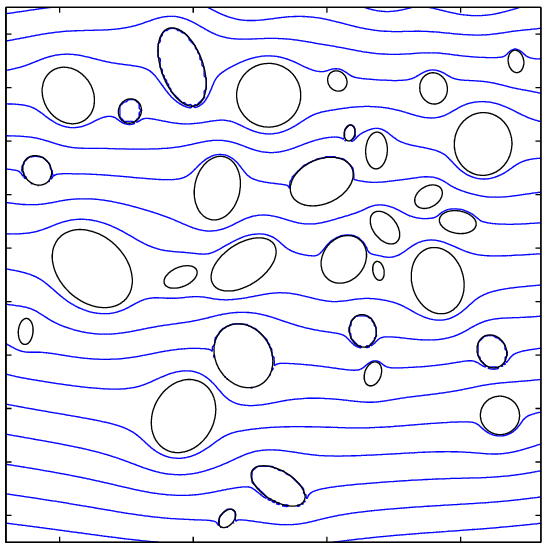}}}
\caption{Stream lines around 300 ellipses. Calculation using the partitioning scheme took 11 minutes on a single 2Ghz PC with a four-core processor.} \label{fig:largeScale}
\end{figure}

We use the partitioning scheme of Section~\ref{sec:large} and set the number of clusters so that each cluster has only a few ellipses. Thus, the run-time and memory requirements for each cluster are kept very moderate. For example, for a geometry with 100 ellipses, we used ten clusters, with the largest cluster having 13 ellipses (not including the ones added as buffers). For each cluster, we used three poles per ellipse (including the ones used as buffers), and stopped the algorithm if the error dropped below $1\%$ or 100 iterations were exhausted. The algorithm was run in parallel on four cores of a desktop PC, using Matlab's parallel computing toolbox, and the run-time for determining the pole and testing point locations was 87s. We then assembled a linear system of size $4832 \times 706$ with the poles and testing-points of all clusters and solved it by QR decomposition in 5s. The error of this solution was $\Delta E =0.013$, slightly higher than the error tolerance for each cluster. The largest geometry solved, which consisted of 300 ellipses, is shown in Fig.~\ref{fig:largeScale}, together with computed flow lines. Here we used 35 clusters, and the run-time for determining the pole and testing points was 9 minutes. The resulting linear system was of size $14581        \times 2150$. Although fairly large, this system has, on average, 7.2 degrees of freedom per curve, and it was solved in 2 minutes by QR decomposition. The error tolerance for each cluster was again $\Delta E < 0.01$ and the error of the full solution was $\Delta E = 0.014$.

\section{Summary}
In this paper, we describe an algorithm for solving 2D Laplace BVPs with a desingularized method. The algorithm is based on rational-function fitting techniques, and it uses rational Gauss-Chebyshev quadrature rules to locate testing points at which the boundary conditions are enforced in a least-squares sense. To use these rules, the boundary curves are parameterized by piecewise-rational functions, which can be used to approximate quite arbitrary boundary curves. The algorithm can be viewed as an adaptive desingularized method, in which the sources are not restricted to curves conformal with the boundary curves, but can be placed anywhere outside the BVP domain. This added flexibility allows the method to adapt to difficult boundary geometries and almost-singular boundary data, yielding solutions with errors approaching the machine epsilon for a relatively small number of degrees of freedom. Examples shown include boundaries with sharp and almost-sharp corners, boundary data that is derived from functions with essential singularities, and boundary curves that nearly touch. We also described a partitioning scheme, in which pole optimization is limited to curves within a cluster, that can be used to tackle large-scale problems.

As the proposed method relies on a correspondence between analytic functions in the complex plane and harmonic functions, and on properties of rational functions, it may be difficult to generalize to 3D Laplace problems. On the other hand, generalizing the method for 2D Helmholtz or Biharmonic BVPs appears more straightforward. The general solution of the Biharmonic equation can be expressed in terms of two functions that are analytic throughout the BVP domain, and rational functions could be used for their approximation. For Helmholz BVPs, it may be possible to exploit a connection between the singularities of the analytic continuations of solution to Helmholtz and Laplace BVPs with the same boundary data. For exterior problems, it was shown in~\cite{millar1980analytic} that the convex hull of singularities of Helmholtz and Laplace BVPs are the same, assuming the same boundary data. Therefore, the location of poles in a Laplace problem can serve as a guide for the location of sources in a Helmholtz problem (see~\cite{leviatan1996analysis} for an example of this idea). In particular, locating the Helmholtz sources so that they enclose the convex hull of singularities of the Laplace problem guarantees that all of the singularities of the Helmholtz problem are enclosed. According to the conjectures in~\cite{barnett2008stability}, which have recently been proved~\cite{kangro2010convergence}, this guarantees exponential convergence.

\section{Acknowledgements}
We would like to acknowledge discussions with Alexandre Megretski, Abe Elfadel, and Stefano Grivet-Talocia, whose seminar on vector fitting led to the approach presented in this paper. We also thank the anonymous reviewers for their insightful comments and suggestions. This work was supported by the Advanced Circuit Research Center at the Technion - Israel Institute of Technology, the Viterbi postdoctoral fellowship, and the Singapore-MIT alliance program in computational engineering.
\appendix
\section{Iterated rational fitting} \label{app:irf}
The IRF method we use is based on a variant of~\cite{coelho1999robust} presented in~\cite{vasilyevequivalence}. As explained in Section~\ref{sec:poleReloc}, we seek a rational function $P/Q$ that approximates $\widehat{W}$, and we do so by solving weighted linear least-squares problems of the from~\eqref{eq:minProbIRF}. Using the quadrature rules of Section~\ref{ssec:quadForIRF}, we discretize the continuous norm in~\eqref{eq:minProbIRF} and write the minimization problem as,
\begin{equation}\label{eq:minProb}
    \min_{\mathbf{a}, \mathbf{b}} \, \sum_{k=1}^K \sum_{n=0}^N{ \dfrac{\sqrt{\lambda_k}}{\widehat{Q}(z_k)}\left|{b_n z_k^n - a_n z_k^n \widehat{W}(z_k) }\right|}
\end{equation}
in which the unknowns are the monomial coefficients of $P(z)$ and $Q(z)$, denoted $\mathbf{b}$ and $\mathbf{a}$, respectively. To avoid the trivial solution, a normalization condition, to be specified below, must be imposed on one of the polynomials.

Let us rewrite~\eqref{eq:minProb} in matrix form, using the notation
\begin{align}\label{eq:notationForIRF}
\mathbf{\Lambda} &= \text{diag}(\lambda_1, \lambda_2, \ldots, \lambda_K),\notag \\
\mathbf{\widehat{Q}} &= \text{diag}\left[\widehat{Q}(z_1), \widehat{Q}(z_2), \ldots, \widehat{Q}(z_K)\right],\notag \\
\mathbf{\widehat{W}} &= \text{diag}\left[\widehat{W}(z_1), \widehat{W}(z_2), \ldots, \widehat{W}(z_K)\right],\\
V_{kn} &= z_k^n, \quad n=0,1,\ldots,N,\notag \\
\mathbf{c} &= \begin{bmatrix}
\mathbf{b}\\
\mathbf{a}\\
\end{bmatrix}.\notag
\end{align}
We obtain,
\begin{equation}\label{eq:irfDiscrete}
\min_{\mathbf{c}} \, \left|\left|\mathbf{\Lambda}^\frac{1}{2} \mathbf{\widehat{Q}}^{-1}\left[\mathbf{V},  -\mathbf{\widehat{W}}\mathbf{V}\right]\mathbf{c}\right|\right|.
\end{equation}
The conditioning of the Vandermonde matrix, $\mathbf{V}$, can be very poor (sampling points on the real line, for example), or very good (sampling points on the unit circle, for example), and this can affect the conditioning of the minimization problem~\eqref{eq:irfDiscrete}. To deal with this problem, we construct two orthonormal bases, one for the column space of $\mathbf{A} = \mathbf{\Lambda}^\frac{1}{2} \mathbf{\widehat{Q}}^{-1}\mathbf{V}$ and one for the column space of $\mathbf{B} = \mathbf{\Lambda}^\frac{1}{2} \mathbf{\widehat{Q}}^{-1}\mathbf{\widehat{W}V}$. This can be done with the Arnoldi iteration~\cite{arnoldi1951principle,saad2003iterative}, by noting that the columns of $\mathbf{A}$ form a Krylov subspace,
\begin{equation}\label{eq:krylovP}
    K= \text{span}[\mathbf{d}, \mathbf{Z}\mathbf{d}, \mathbf{Z}^2\mathbf{d}, \ldots, \mathbf{Z}^N\mathbf{d}],
\end{equation}
where $\mathbf{Z} = \mathrm{diag}\left(z_1, z_2, \ldots, z_K\right)$ and $\mathbf{d} = \mathrm{diag}(\mathbf{\Lambda}^\frac{1}{2} \mathbf{\widehat{Q}}^{-1})$, and similarly for the columns of $\mathbf{B}$. Once we have constructed the orthonormal bases,
\begin{align}
\mathrm{colsp}(\mathbf{P}) &= \mathrm{colsp}(\mathbf{A}), \quad \mathbf{P^\dagger P} = \mathbf{I} \\
\mathrm{colsp}(\mathbf{Q}) &= \mathrm{colsp}(\mathbf{B}), \quad \mathbf{Q^\dagger Q} = \mathbf{I},
\end{align}
we rewrite~\eqref{eq:irfDiscrete} in the new basis,
\begin{equation}\label{eq:irfDiscretePQ}
\min_{\mathbf{\tilde{c}}} \, \left|\left|\left[\mathbf{P}, -\mathbf{Q}\right]\mathbf{\tilde{c}}\right|\right|,
\end{equation}
and impose the normalization condition $\tilde{c}_{2N} = 1$, which corresponds to $Q(z)$ being monic in the new basis. The solution of the minimization problem is then obtained by solving the over-determined linear system
\begin{equation}\label{eq:irfLS}
[\mathbf{P},-\mathbf{Q}_{1:N-1}]\mathbf{\tilde{c}} = \mathbf{Q}_N
\end{equation}
in the least-squares sense. In~\eqref{eq:irfLS}, we denote the column corresponding to the $N$th-degree polynomial basis function by $\mathbf{Q}_{N}$, while the rest of the $\mathbf{Q}$ matrix is denoted by $\mathbf{Q}_{1:N-1}$.
The numerical stability of~\eqref{eq:irfDiscretePQ} is generally much better than that of~\eqref{eq:irfDiscrete}, as the unknowns are coefficients of two sets of orthonormal columns. 

Once~\eqref{eq:irfDiscrete} is solved, the zeros of $Q(z)$, which are the poles for the next iteration, can be obtained without transforming back to the monomial basis. They are the eigenvalues of a generalized companion matrix~\cite{coelho1999robust},
\begin{equation}\label{eq:roots}
    \mathbf{C} = \mathbf{H}_{1:N,1:N} - H_{N+1,N}\,\mathbf{\tilde{c}}_{N+1:2N}\,\mathbf{e}_N^T,
\end{equation}
where $\mathbf{e}_N$ is the $N^\text{th}$ column of the order-$N$ identity matrix and $\mathbf{H}$ is the upper Hessenberg matrix obtained from the Arnoldi iteration used to generate the orthonormal basis for colsp($\mathbf{B}$).

\section{Determining whether a pole is inside $D$} \label{app:in_or_out}
When a pole is close to the boundary, or the boundary shape is complicated, it may be nontrivial to find out whether a pole is inside $D$ or not. By restricting ourselves to boundaries described by piecewise rational functions, a solution to this problem is readily obtained. Using Cauchy's theorem, a pole $z'$ is inside $D$ if and only if
\begin{equation}\label{eq:isIn}
 \int_C \dfrac{1}{z-z'}dz =  \int_0^1 \dfrac{1}{z(s)-z'}\dfrac{dz(s)}{ds} \,ds = 2\pi i
\end{equation}
This integral cannot be calculated exactly using the rational Gauss-Chebyshev rules because the Chebyshev weight function is absent. However, it can be calculated analytically by breaking it up into rational-function sections, writing the integrands as rational functions, and using a partial-fraction expansion.

\clearpage
\bibliographystyle{elsarticle-num}
\bibliography{laplace}
\end{document}